\documentclass[12pt,hidelinks]{article}
\usepackage{mathrsfs,amsmath,epsfig,amssymb,color,graphicx,bm}
\usepackage{algorithm,algorithmic,subcaption,hyperref}
\usepackage{hyperref}
\usepackage[round, sort]{natbib}
\usepackage[margin=1in]{geometry}
\usepackage[symbol]{footmisc}
\usepackage{bm} 
\usepackage[figuresright]{rotating}
\usepackage{rotating}
\usepackage{amsmath}
\usepackage{arydshln}
\usepackage{indentfirst}  
\usepackage{palatino}
\usepackage{tikz}
\usepackage{diagbox}
\usepackage{longtable}
\usepackage{booktabs}
\usepackage{hyperref} 
\usepackage{multirow}
\usepackage{pifont}
\usepackage[section]{placeins} 
\usepackage{threeparttable}
\usepackage[hyphenbreaks]{breakurl} 
\usetikzlibrary{shapes.geometric, arrows}
\graphicspath{{figures/}}
\allowdisplaybreaks

\begin{document}
\captionsetup[figure]{labelfont={bf},name={Figure}, labelsep=period}
	
\centerline {\Large  \bf An Online Updating Approach for Estimating and  Testing   }
\centerline {\Large\bf Mediation Effects  with  Big Data Streams }
\vspace*{0.2in}

\centerline{ { Xueyan Bai and Haixiang Zhang$^{*}$}}
\vspace*{0.1in}

\centerline{\it \small Center for Applied Mathematics, Tianjin University, Tianjin 300072, China}

\footnotetext[1]{Corresponding author: haixiang.zhang@tju.edu.cn (Haixiang Zhang)}
\vspace{1cm}

\begin{abstract}
\noindent The use of mediation analysis has become increasingly popular in various research fields in recent years. The primary objective of mediation analysis is to examine the indirect effects along the pathways from exposure to outcome.
Meanwhile, the advent of data collection technology has sparked a surge of interest in the burgeoning field of big data analysis, where mediation analysis of streaming data sets has recently garnered significant attention. The enormity of the data, however, results in an augmented computational burden.
The present study proposes an online updating approach to address this issue, aiming to estimate and test mediation effects in the context of linear and logistic mediation models with massive data streams. The proposed algorithm significantly enhances the computational efficiency of Sobel test, adjusted Sobel test, joint significance test, and adjusted joint significance test.
We conduct a substantial number of numerical simulations to evaluate the performance of the renewable method. Two real-world examples are employed to showcase the practical applicability of this approach.
\end{abstract}

\section{Introduction}
The analysis of mediation plays a crucial role in comprehending the causal mechanism by which the independent variable $X$ influences the dependent variable $Y$ through the intermediate variable $M$ \cite[]{MacKinnon2008-Book}.  The topic of mediation analysis has been extensively explored across diverse disciplines, such as psychology \citep{physical}, economics \citep{economics},  medicine \citep{medicine}, sociology \citep{sociology}, and behavioral science \citep{behavioral_science}. 
After the establishment of the  statistical formula for mediation analysis by \cite{Baron}, there has been a remarkable surge in both methodological advancements and practical applications of mediation methods. For example, \cite{FMA-JASA-2012} studied the functional causal mediation analysis with an application to brain connectivity; \cite{Rob-MA-2014} explored the application of robust mediation analysis using median regression; \cite{Zhangetal2016-HIMA} proposed a novel framework for high-dimensional linear mediation analysis with sure independent screening and minimax concave penalty techniques;  \cite{Compositional-AOAS-2019} proposed a sparse compositional mediation model that can be used to estimate the causal direct and indirect effects; 
\cite{zhang-surHIMA-2021} considered mediation analysis for survival data with high-dimensional mediators; \cite{Bayesian-SEM-2021} studied Bayesian causal mediation analysis with latent mediators and survival outcome; \cite{SEM-MacKinnon-2021} demonstrated the similarities and differences between the causal and traditional estimators for mediation models; \cite{qHIMA-2024} proposed a novel approach for selecting and testing mediators in the framework of high-dimensional quantile mediation analysis.

In the era of big data, streaming datasets, which are continuously generated and received in real-time, have become prevalent in various domains such as investment analysis, medical imaging, and computer vision. The analysis of streaming data is increasingly indispensable and valuable due to its ability to handle large volumes of data that arrive sequentially. Given the limited memory capacity that can only accommodate a small batch of data at a time, efficient analysis without relying on historical data becomes imperative. The recent literature has presented advanced statistical methodologies for handling big data streams. e.g., \cite{Lou_2019_Renewable} presented an incremental updating algorithm to analyse streaming data sets using generalized linear models; \cite{online-cox-JCGS} developed an online updating method for carrying out survival analysis under the Cox model;  \cite{streamed-LD-2023} proposed a novel estimation and inference framework for dynamically updating point estimates and their corresponding standard errors based on sequentially collected datasets; \cite{Luo-Zhou-S-JASA-2023}  developed an incremental learning algorithm to analyze streaming datasets with correlated outcomes;  \cite{quantile-JOE-2024} developed a sequential algorithm to support efficient quantile regression  analysis for stream data; \cite{multiplicative-2024} proposed a renewable learning method for the multiplicative regression model with streaming data.

The current findings on mediation analysis in the context of big data streams are limited.  
In this study, we propose an online updating approach to tackle this issue, with the aim of estimating and testing mediation effects within the framework of linear and logistic mediation models amidst vast data streams. The proposed algorithm significantly amplifies the computational efficiency of Sobel test \cite[]{Sobel-1982}, adjusted Sobel test \cite[]{AJS-2024}, joint significance test \cite[]{David-PM-2002}, and adjusted joint significance test \cite[]{AJS-2024}. The primary advantage of our approach lies in its utilization of newly arrived data batches, in conjunction with summary statistics derived from historical raw data. This not only ensures computational efficiency but also minimizes storage requirements during the process of conducting mediation analysis for large-scale data streams.

The remainder of this paper is organized as follows.  In Section 2, we provide a comprehensive overview of linear mediation models. Additionally, we propose an  online updating approach for testing mediation effects utilizing four distinct methods: the Sobel test, adjusted Sobel test, joint significance test, and adjusted joint significance test. In Section 3, we give an online updating approach for logistic mediation models with large-scale data streams. The performance of the proposed method is evaluated through simulations in Section 4. Two real data examples are provided in Section 5. The paper concludes with some remarks in Section 6.

\textsl{}
\section{Linear Mediation Model}\label{sec-2}
\setcounter{equation}{0}
\subsection{Online Updating Estimation}

In this section, we provide a comprehensive overview of the counterfactual (or potential outcome) framework for linear mediation models that involve multiple continuous mediators and a continuous outcome variable.

\begin{figure}[H]
\tikzstyle{startstop} = [rectangle,  minimum width = 1cm, minimum height=1cm,text centered, draw = black]
\tikzstyle{process} = [rectangle, minimum width=1cm, minimum height=1cm, text centered, draw=black]	
\tikzstyle{point}=[coordinate]
\tikzstyle{arrow} = [->,>=stealth]
\tikzstyle{line} =[-]
	
	\begin{center}
		
	\begin{tikzpicture}[node distance=2cm]
\node[startstop](X){$X$};
\node[process,right of = X, yshift = 3cm, xshift = 4cm](M1){$M_1$};
\node[process,right of = X, yshift = 1.5cm, xshift = 4cm](M2){$M_2$};
\node[process,right of = X, yshift = -2cm, xshift = 4cm](Mp-1){$M_{p-1}$};
\node[process,right of = X, yshift = -4cm, xshift = 4cm](Mp){$M_p$};
\node[process,right of = X, xshift = 10cm](Y){$Y$};

\draw [arrow] (X) -- node [yshift=0.5cm] {$\alpha_1$}(M1);
\draw [arrow] (X) -- node [yshift=0.3cm, xshift=0.2cm] {$\alpha_2$}(M2);
\draw [arrow] (X) -- node [yshift=0.2cm] {$\alpha_{p-1}$}(Mp-1);
\draw [arrow] (X) -- node [yshift=0.2cm] {$\alpha_p$}(Mp);
\draw (6,0.55) circle (0.03cm) [fill= black];
\draw (6,0.35) circle (0.03cm) [fill= black];
\draw (6,0.15) circle (0.03cm) [fill= black];
\draw (6,-0.15) circle (0.03cm) [fill= black];
\draw (6,-0.35) circle (0.03cm) [fill= black];
\draw (6,-0.55) circle (0.03cm) [fill= black];
\draw [arrow] (X) --  node [yshift=0.3cm, xshift=0.5cm] {$\gamma$}(Y);
\draw [arrow] (M1) -- node [yshift=0.3cm, xshift=0cm] {$\beta_1$}(Y);
\draw [arrow] (M2) -- node [yshift=0.3cm, xshift=-0.4cm] {$\beta_2$}(Y);
\draw [arrow] (Mp-1) -- node [yshift=0.2cm, xshift=-.3cm] {$\beta_{p-1}$}(Y);
\draw [arrow] (Mp) -- node [yshift=0.2cm, xshift=-.3cm] {$\beta_p$}(Y);

	\end{tikzpicture}
\caption{The scenario involving multiple mediators of mediation model.}
\label{fig-1-1}
\end{center}
\end{figure}
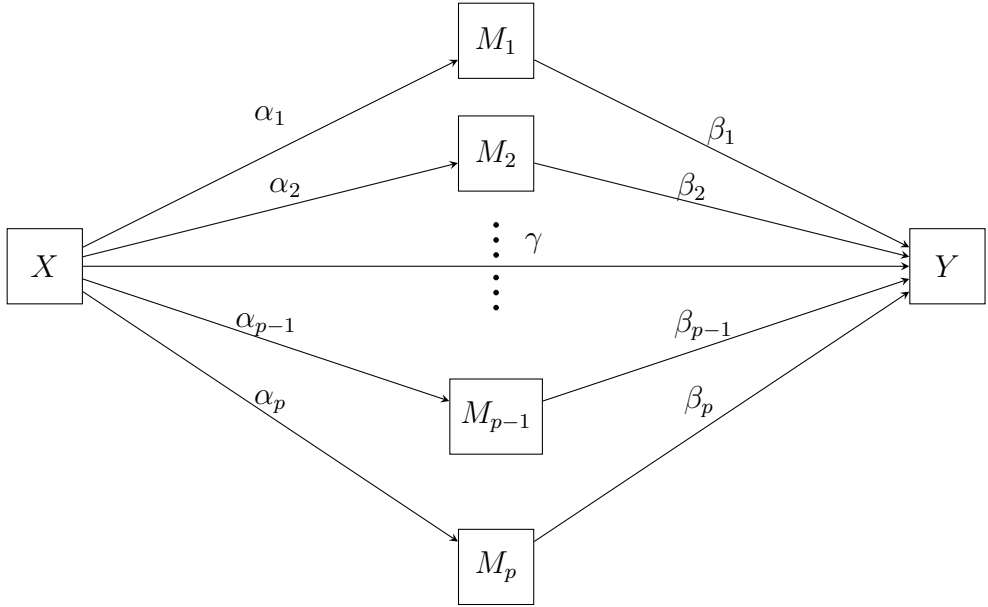

Based on the notations of counterfactuals
\citep{2009_ConceptualIC}, we denote  $\boldsymbol{M}(x)=(M_{1}(x),\ldots, M_{p}(x))^{'}$ 
as the potential value of a p-dimensional vector of mediators under exposure level $x$.
Denote $Y(x,\boldsymbol{m})$
as the potential outcome when $x$ is the exposure
and $ \boldsymbol{m} = (m_1,\ldots,m_p)^{'}$	are the mediators.
The linear mediation models can be represented as follows (Figure \ref{fig-1-1}):
\begin{align}
Y(x,\boldsymbol{m}) &=  \gamma x +
\beta_{1} m_1 + \cdots + \beta_{p} m_p
+ \bm{\theta}^{'}  \bm{Z}
+ \epsilon,
\label{EQ-2-1} \\
M_{j}(x) &=  \alpha_{j} x + \bm{\eta}_{j}^{'} \bm{Z} + e_{j}, j=1,\ldots, p,
\label{EQ-2-2}
\end{align}
where $\boldsymbol{Z}= (Z_1,\ldots,Z_q)^{'}$ is the vector of confounders; $\epsilon$ and $\bm{e}=(e_1,\ldots,e_p)^{'}$ are the error terms; $\boldsymbol{\alpha} = (\alpha_1,\ldots,\alpha_p)^{'}$ is a vector of regression coefficients between the exposure and mediators after adjusting $\bm{Z}$, and
$\boldsymbol{\beta} = (\beta_1,\cdots,\beta_p)^{'}$  is the vector of regression coefficients between mediators and outcome after adjusting $\bm{Z}$; $E(\epsilon) = E(e_j) = 0$ for $j=1,\ldots,p$.


For the purpose of identifying natural direct efect 
(NDE) and natural indirect efects (NIE), the causal
mediation analysis  requires  four
fundamental assumptions \citep{2009_ConceptualIC}:\\
(C.1) Stable Unit Treatment Value Assumption (SUTVA).
There is no multiple versions of exposures and there
is no interference between subjects, which implies
that the observed variables are identical to the potential variables corresponding to the actually observed exposure level. i.e.,
$\bm{M}= \sum\limits_{x} \bm{M}(x) \bm{I}(X = x)$, and $Y= \sum\limits_{x} \sum\limits_{\bm{m}} Y(x,\bm{m}) \bm{I}(X = x, \bm{M} = \bm{m})$
, where
$\bm{I}(\cdot)$is the indicator function.
\\
(C.2)There are no measurement errors in the mediators and
the outcome.
\\
(C.3)
Sequential ignorability:\\
\hspace*{1cm}(i) No unmeasured confounding between exposure and
the potential outcome: $Y(x,\bm{m}) \perp X \mid \bm{Z}$.
\\
\hspace*{1cm}(ii)
No unmeasured confounding for the mediator-
outcome relationship after adjusting for the exposure:
$Y(x,\bm{m}) \perp \bm{M} \mid \{X,\bm{Z}\}$.
\\
\hspace*{1cm}(iii)
No unmeasured confounding for the exposure effect on all the mediators:
$\bm{M}(x) \perp X \mid \bm{Z}$.
\\
\hspace*{1cm}(iv)
No exposure-induced confounding between mediators and the potential outcome:
$Y(x,\bm{m}) \perp \bm{M}(x^{*}) \mid \bm{Z}$
\\
(C.4) The mediators are assumed to be causally independent. i.e.,  one mediator is not the cause of another.
 
 Under the above four assumptions, we denote the effect of an exposure on outcome as the total effect (TE), which can be decomposed into 
NDE and  NIE
towards models \eqref{EQ-2-1} and \eqref{EQ-2-2}. Based on Theorem 1 of \cite{big-med-2023}, we know that
 \begin{align*}
 NDE &= \gamma (x -x^*),\\
 NIE &= \sum\limits_{j =1}^{p} \alpha_{j} \beta_{j} (x - x^*),\\
 TE &= \Big(
 \gamma + \sum\limits_{j =1}^{p} \alpha_{j} \beta_{j}
 \Big)(x - x^*).
 \end{align*}
Therefore, the term $\alpha_k\beta_k$ can be interpreted as the causal indirect effect transmitted by the kth mediator $M_k$ along the pathway $X \rightarrow M_k \rightarrow Y$ (see Figure \ref{fig-1-1}).

The focus of this paper is on the analysis of scenery with streaming data sets, specifically aiming to estimate and test mediation effects denoted by $\alpha_k\beta_k$'s. By \cite{Lou_2019_Renewable}, the key concept is to employ a Taylor expansion on the loss function in order to avoid the necessity of using historical individual-level data. The statistics obtained from Taylor expansions can be employed to succinctly summarize the information contained in historical data. To be specific, we suppose that the data streams are represented as
 $ \boldsymbol{D_{1}} =\{\boldsymbol{Y}^{(1)}, \boldsymbol{W}^{(1)} \}, 
 \boldsymbol{D_{2}} =\{\boldsymbol{Y}^{(2)}, \boldsymbol{W}^{(2)} \}, \cdots , 
 \boldsymbol{D_{k}} =\{\boldsymbol{Y}^{(k)}, \boldsymbol{W}^{(k)} \}$, 
where $ \boldsymbol{Y}^{(i)}  = \big( Y_{1}^{(i)}, \ldots , Y_{n_i}^{(i)}  \big)^{\bm{'}}$ denotes the outcomes from the $i$th batch with mean $\boldsymbol{\mu}^{(i)}  = \big( \mu_{1}^{(i)},\ldots, \mu_{n_i}^{(i)} \big) ^{\bm{'}}$; 
$\boldsymbol{ W}^{(i)} \in \mathbb{R}^{n_i \times (1+p+q)}$ is a matrix consisting of exposures, mediators and covariates,
and 
\begin{align}
\label{EQ-2-3}
\boldsymbol{W}^{(i)} =\left[ \begin{array}{cccccccc}
X_{{1}}^{(i)} & M_{{11}}^{(i)} & M_{{12}}^{(i)} & \cdots & M_{{1p}}^{(i)} &  Z_{{11}}^{(i)} & \cdots & Z_{{1q}}^{(i)} \\
X_{{2}}^{(i)} & M_{{21}}^{(i)} & M_{{22}}^{(i)} & \cdots & M_{{2p}}^{(i)} &  Z_{{21}}^{(i)} & \cdots & Z_{{2q}}^{(i)} \\
\cdots & \cdots & \cdots & \cdots & \cdots & \cdots& \cdots & \cdots\\ 
X_{{n_i}}^{(i)} & M_{{n_i 1}}^{(i)} & M_{{n_i 2}}^{(i)} & \cdots & M_{{n_i p}}^{(i)} &  Z_{{n_i 1}}^{(i)} & \cdots & Z_{{n_i q}}^{(i)} \\
\end{array}\right].
\end{align}
Denote $N_k$ as the total sample size of aggregated streaming data up to batch $k$, where $N_k = \sum \limits_{i =1}^{k} n_i$, and $n_i$ is the size of $i$th data batch. For $i=1,\cdots,k$, 
the linear model \eqref{EQ-2-1} with the $i$th batch $\boldsymbol{D}_i$ can be reformulated as 
\begin{align*}
Y_{1}^{(i)} &= \gamma X_{{1}}^{(i)} + \beta_{1} M_{{11}}^{(i)} + \beta_{2} M_{{12}}^{(i)} + \cdots + \beta_{p} M_{{1p}}^{(i)} + \theta_{1} Z_{{11}}^{(i)} + \cdots + \theta_{q} Z_{{1q}}^{(i)} + \epsilon_{1}^{(i)}, \\
\notag Y_{2}^{(i)} &=\gamma X_{{2}}^{(i)} + \beta_{1} M_{{21}}^{(i)} + \beta_{2} M_{{22}}^{(i)} + \cdots + \beta_{p} M_{{2p}}^{(i)} + \theta_{1} Z_{{21}}^{(i)} + \cdots + \theta_{q} Z_{{2q}}^{(i)} + \epsilon_{2}^{(i)},\\
\cdots 
\\
\notag Y_{n_i}^{(i)} &= \gamma X_{{n_i}}^{(i)} + \beta_{1} M_{{n_i1}}^{(i)} + \beta_{2} M_{{n_i2}}^{(i)} + \cdots + \beta_{p} M_{{n_ip}}^{(i)} + \theta_{1} Z_{{n_i1}}^{(i)} + \cdots + \theta_{q} Z_{{n_iq}}^{(i)} + \epsilon_{n_i}^{(i)},
\end{align*}
where the notations have analogous connotations to those of model \eqref{EQ-2-1}.
For convenience,
we denote  $\boldsymbol{\epsilon}^{(i)} = \big(\epsilon_{1}^{(i)}, \ldots ,\epsilon_{n_i}^{(i)} \big)^{\bm{'}}$ and
$\bm{\Gamma} = \left( \gamma, \beta_{1}, \ldots,\beta_{p}, \theta_{1}, \ldots \theta_{q} \right)^{\boldsymbol{'}}  $.
The linear model \eqref{EQ-2-1} can be represented in matrix form for the $i$th batch $\boldsymbol{D}_i$ as follows:
\begin{align}
\label{EQ-2-4} 
\boldsymbol{ Y}^{(i)}  =\boldsymbol{ W}^{(i)}  \boldsymbol{ \Gamma } + \boldsymbol{\epsilon }^{(i)}, \quad i=1, \ldots,k.
\end{align}
The tilde is employed to denote a symbol representing a sequentially updated quantity.
For instance, the symbol $\boldsymbol{ \tilde{\Gamma}}^{(i)} $ represents a renewable estimator, which is initialized by the estimator obtained from the first batch of data $\boldsymbol{D_1}$. The previous estimator $\boldsymbol{ \tilde{\Gamma}}^{(i-1)}$ is sequentially updated to $\boldsymbol{ \tilde{\Gamma}}^{(i)} $ as each subsequent data batch $\boldsymbol{D_i}$ arrives.
The objective here is to derive a renewable estimator, along with its corresponding standard error, for the $\boldsymbol{{\Gamma}}$.

 

Denote $\boldsymbol{U}^{(i)}$ as the score function with batch $\boldsymbol{D_{i}}$, which is given by
 $\boldsymbol{U}^{(i)}  = (\boldsymbol{W}^{(i)})^{\boldsymbol{'}} (\boldsymbol{Y}^{(i)}
  - \boldsymbol{W}^{(i)}\boldsymbol{\Gamma}) $.
The corresponding negative Hessian matrix of  $\boldsymbol{U}^{(i)}$  is 
$\boldsymbol{J}^{(i)}
=(\boldsymbol{W}^{(i)})^{\boldsymbol{'}}  \boldsymbol{W}^{(i)}$.
Let $\boldsymbol{\tilde{J}}^{(k)} = \sum_{i=1}^{k} \boldsymbol{J^{(i)}}$
be the sum of $k$ negative Hessian matrices.
Based on \cite{Lou_2019_Renewable}, 
we obtain the renewable estimator $\bm{\tilde{\Gamma}}^{(k)} $ when the $k$th data batch is available:
\begin{align*}
\boldsymbol{\tilde{\Gamma }}^{(k)} &= 
\big( \boldsymbol{\tilde{J}}^{(k-1)}  + \boldsymbol{J}^{(k)} \big)^{-1} 
\big\{ \boldsymbol{\tilde{J}}^{(k-1)} \boldsymbol{\tilde{\Gamma }}^{(k-1)} + (\boldsymbol{W}^{(k)})^{\boldsymbol{'}} \boldsymbol{Y}^{(k)} \big\},
\end{align*}
where $\bm{\tilde{\Gamma}}^{(0)} = \boldsymbol{0}_{}$ and  $\boldsymbol{\tilde{J}}^{(0)}= \boldsymbol{0}_{}$. The renewable estimator of parameter $\boldsymbol{\beta}$ is
\begin{align}\label{betahat}
\boldsymbol{ \tilde{\beta}}^{(k)}  = (\tilde{\beta}_{1}^{(k)}, \tilde{\beta}_{2}^{(k)}, \ldots , \tilde{\beta}_{p}^{(k)})
= \big( \boldsymbol{ \tilde{\Gamma}}_{2}^{(k)} ,\boldsymbol{ \tilde{\Gamma} }_{3}^{(k)}, \ldots, \boldsymbol{ \tilde{\Gamma}}_{p+1 }^{(k)}   \big),
\end{align}
where $\boldsymbol{ \tilde{\Gamma}}_{j}^{(k)} $stands for the $j$th item of $\boldsymbol{ \tilde{\Gamma}}^{(k)} $ for $j=2,\ldots, p+1$. The details on a renewable estimator for $Var(\epsilon)$ are provided in the Appendix, which is expressed as follows:
\begin{align}
\label{A1}
 \tilde{\phi}^{(k)} & = \dfrac{1}{N_k - (1+p+q)} \sum\limits_{i=1}^{k} \big(\boldsymbol{Y}^{(i)} - \boldsymbol{W}^{(i)}  \boldsymbol{\tilde{\Gamma}}^{(k)}  \big)^{\bm{'}} 
\big(\boldsymbol{Y}^{(i)}  - \boldsymbol{W}^{(i)}  \boldsymbol{\tilde{\Gamma}}^{(k)}  \big)\\
\nonumber & = \dfrac{1}{N_k - (1+p+q)} \{ \big(N_{k-1} - (1+p+q)  \big)\tilde{\phi}^{(k-1)}  + (\boldsymbol{\tilde{\Gamma}}^{(k-1)})^{\bm{'}} \boldsymbol{\tilde{J}}^{(k-1)}  \boldsymbol{\tilde{\Gamma}}^{(k-1)}  \\
\nonumber & + 
(\boldsymbol{Y}^{(k)}){'}\boldsymbol{ Y}^{(k)}  -(\boldsymbol{\tilde{\Gamma}}^{(k)})^{\bm{'}} \boldsymbol{\tilde{J}}^{(k)} \boldsymbol{\tilde{\Gamma}}^{(k)}  \}.
\end{align}
From \cite{Lou_2019_Renewable}, the estimated variance of $\boldsymbol{\tilde{\Gamma}}^{(k)}$ is
$\boldsymbol{\Sigma_{{\tilde{\Gamma}}^{(k)}}} = \tilde{\phi}^{(k)}
\big(\boldsymbol{\tilde{J}}^{(k-1)} + \boldsymbol{J}^{(k)} \big)^{-1} $.
By taking the diagonal of $\boldsymbol{\Sigma_{\tilde{\bm{\Gamma}}^{(k)}}}$,  the estimated standard variance of ${\tilde{\beta }_j}^{(k)}$ is 
\begin{eqnarray}\label{betaSE}
({\tilde{\sigma}_{\beta_1}^{(k)}},\ldots,{\tilde{\sigma}_{\beta_p}^{(k)}}) =
\big( \sqrt{(\boldsymbol{\Sigma_{\tilde{\bm{\Gamma}}^{(k)}}})_{22}}, \ldots, \sqrt{(\boldsymbol{\Sigma_{\tilde{\bm{\Gamma}}^{(k)}}})_{p+1,p+1}}  \big). 
\end{eqnarray}

Next, the renewable estimator for $\boldsymbol{\alpha}$, which characterizes the effects along the pathway $X \rightarrow M_k$, is taken into consideration. For the ith data batch $\bm{D_i}$, 
we denote the mediator matrix as 
\begin{eqnarray*}
\boldsymbol{M^{(i)}}&=&\left[ \begin{array}{cccc}
M_{{11}}^{(i)} & M_{{12}}^{(i)} &\cdots & M_{{1p}}^{(i)}\\
M_{{21}}^{(i)} & M_{{22}}^{(i)} &\cdots & M_{{2p}}^{(i)}\\
\cdots & \cdots & \cdots & \cdots\\ 
M_{{n_i 1}}^{(i)} &M_{{n_i 2}}^{(i)} & \cdots & M_{{n_i p}}^{(i)}\\ 
\end{array}\right]\\
&=&\big( \boldsymbol{\omega}_{1}^{(i)}, \boldsymbol{\omega}_{2}^{(i)}, \cdots, \boldsymbol{\omega}_{p}^{(i)} \big),
\end{eqnarray*}
where $\boldsymbol{\omega}_{j}^{(i)} \in \mathbb{R}^{n_i}$ is the $j$th column of $\boldsymbol{M^{(i)}}$, $i=1,\ldots k$. In a similar way, let

\begin{eqnarray*}
\boldsymbol{S^{(i)}}=\left[ \begin{array}{cccc}
X_{{1}}^{(i)} & Z_{{11}}^{(i)} & \cdots & Z_{{1q}}^{(i)}\\
X_{{2}}^{(i)} & Z_{{21}}^{(i)} & \cdots & Z_{{2q}}^{(i)}\\
\cdots & \cdots & \cdots & \cdots\\ 
X_{{n_i}}^{(i)} & Z_{{n_i 1}}^{(i)} & \cdots & Z_{{n_i q}}^{(i)}\\ 
\end{array}\right].
\end{eqnarray*}
The linear model of \eqref{EQ-2-2} with the ith data batch can be written as 
\begin{align*}
M_{1j}^{(i)} & = {\alpha}_{j}^{} X_{1}^{(i)}  + {\eta}_{1}^{} Z_{11}^{(i)} + \cdots + {\eta}_{q}^{} Z_{1q}^{(i)} +e_{1}^{(i)},
\\
 M_{2j}^{(i)} & =  {\alpha}_{j}^{} X_{2}^{(i)}  + {\eta}_{1}^{} Z_{21}^{(i)} + \cdots + {\eta}_{q}^{} Z_{2q}^{(i)} + e_{2}^{(i)},
 \\
 \cdots
 \\
  M_{n_i j}^{(i)} &=  {\alpha}_{j}^{} X_{n_{i} }^{(i)}  + {\eta}_{1}^{} Z_{n_i 1}^{(i)} + \cdots + {\eta}_{q}^{} Z_{n_i q}^{(i)} + e_{n_i}^{(i)}.
\end{align*}
Let $\bm{e^{(i)}} = \big( e_{1}^{(i)}, \ldots , e_{n_i}^{(i)} \big)^{\boldsymbol{'}}$, then we have the following matrix expression:
\begin{align}
\label{EQ-2-5}
\boldsymbol{\omega} _{j}^{(i)}=  \boldsymbol{S}^{(i)} 
\boldsymbol{{\lambda}}_{j}^{} +\boldsymbol{e}^{(i)}, \quad j= 1,\ldots,p,
\end{align} 
where 
$\boldsymbol{{\lambda}}_{j}^{}
= ({\alpha}_{j}^{}, \eta_1,\cdots,\eta_q)^{\bm{'}}$.
Similarly, the score function with the $i$th data batch $\bm{D_i}$ is
\begin{align*}
\boldsymbol{V}^{(i)} = (\boldsymbol{S}^{(i)})^{'} \big( \boldsymbol{\omega}_{j}^{(i)} - \boldsymbol{S}^{(i)} \boldsymbol{\lambda}_{j}^{}
\big), ~ j= 1, \ldots, p.
\end{align*}
The negative Hessian matrix of 
$\boldsymbol{V}^{(i)}$ is calculated as 
$\boldsymbol{H}^{(i)} 
= (\boldsymbol{S}^{(i)})^{'} \boldsymbol{S^{(i)}}$.
We denote $\boldsymbol{\tilde{H}}^{(k)} = \sum_{i =1}^{k} \boldsymbol{H}^{(i)}$
as the sum of $k$ negative Hessian matrices. 
A renewable expression for $\boldsymbol{\tilde{\lambda}}_{j}^{(k)}$  is denoted as
\begin{align*}
\boldsymbol{\tilde{\lambda}}_{j}^{(k)}
 = \big( 
 \boldsymbol{\tilde{H}}^{(k-1)} +
  \boldsymbol{H}^{(k)} \big)^{-1}
\big( \boldsymbol{\tilde{H}}^{(k-1)}   \boldsymbol{\tilde{\lambda}}_{j}^{(k-1)}+ (\boldsymbol{S}^{(k)})^{'} \boldsymbol{\omega}_{j}^{(k)} \big).
\end{align*}
When the $k$th data batch $\boldsymbol{D}_k$ is avaliable,
we get a renewable estimator for $\boldsymbol{\alpha}$
\begin{align}
\boldsymbol{\tilde{\alpha}}^{(k)} = (\tilde{\alpha}_{1}^{(k)},\ldots, \tilde{\alpha}_{p}^{(k)} )^{\bm{'}}
= \left( (\boldsymbol{\tilde{\lambda}}_{1}^{(k)})_1,
\ldots, (\boldsymbol{\tilde{\lambda}}_{p}^{(k)})_1  \right)^{\bm{'}},
\label{EQ-2-6} 
\end{align}
where $(\boldsymbol{\tilde{\lambda}}_{j}^{(k)})_1$
is the first element of $\boldsymbol{\tilde{\lambda}}_{j}^{(k)}$ for $ j=1,\ldots,p$.
The variance matrix of $\boldsymbol{\tilde{\lambda}}_{j}^{(k)}$  is denoted as
$\boldsymbol{\Sigma_{\tilde{\lambda}_{j}^{(k)}}} =\tilde{\varphi}_{j}^{(k)} \big( \boldsymbol{\tilde{H}}^{(k-1)} + \boldsymbol{H}^{(k)} \big)^{-1}$,
where
\begin{align*}
\tilde{\varphi}_{j}^{(k)} & = \dfrac{1}{N_{k} - (1+q)}
\sum\limits_{j =1}^{k} \big( \boldsymbol{\omega}_{j}^{(k)}  - \boldsymbol{S}^{(j)} \boldsymbol{\tilde{\lambda}}_{j}^{(k)} \big)^{\bm{'}}
\big( \boldsymbol{\omega}_{j}^{(k)} - \boldsymbol{S}^{(j)} \boldsymbol{\tilde{\lambda}}_{j}^{(k)} \big)\\
& = \dfrac{1}{N_{k} - (1+q)}
\{(N_{k-1} - (1+q)) 
\tilde{\varphi}_{j}^{(k-1)} +
\boldsymbol{(\tilde{\lambda}}_{j}^{(k)})^{\bm{'}}
\boldsymbol{\tilde{H}}^{(k-1)}  \boldsymbol{\tilde{\lambda}}_{j}^{(k)} 
+ (\boldsymbol{ \omega}_{j}^{(k)})^{\bm{'}} \boldsymbol{\omega}_{j}^{(k)} 
- (\boldsymbol{\tilde{\lambda}}_{j}^{(k)})^{\bm{'}}
\boldsymbol{\tilde{H}}^{(k)} \boldsymbol{\tilde{\lambda}}_{j}^{(k)}
\}.
\end{align*}
Denote  
$\big( \boldsymbol{\Sigma_{\tilde{\lambda}_{j}^{(k)}}} \big) _{11}$ as the first diagonal element of the variance matrix $\boldsymbol{\Sigma_{\tilde{\lambda}_{j}^{(k)}}}$, the estimated standard error of $\boldsymbol{\tilde{\alpha}}^{(k)}$ is 
\begin{align}
({\tilde{\sigma}_{\alpha_1}^{(k)}}, \ldots, {\tilde{\sigma}_{\alpha_p}^{(k)}}) = \left( \sqrt{( \boldsymbol{\Sigma_{\tilde{\lambda}_{1}^{(k)}}})_{11}},
\ldots,
\sqrt{(\boldsymbol{\Sigma_{\tilde{\lambda}_{p}^{(k)}}})_{11}} \right)^{}.
\label{EQ-2-7}
\end{align}

\subsection{Multiple Testing and Confidence Interval with Streaming Data}\label{sec22}
The current focus of our study lies in the hypothesis testing for mediation effects within the context of streaming data. Let $\Omega = \{j: \alpha_j\beta_j\neq 0, j=1,\cdots,p\}$ be the index set of significant (or active) mediators. Under the significance level $\delta$, we consider the following multiple testing problem:
\begin{align}\label{MT-1}
H_{0j}: \alpha_j\beta_j = 0 \leftrightarrow H_{Aj}: \alpha_j\beta_j \neq 0, ~j=1,\ldots,p,
\end{align}
where  each null hypothesis $H^{}_{0j}$ is composite with three components:
\begin{eqnarray*}
&&H^{}_{00,j}: \alpha_j=0, \beta_j=0;\\
&&H^{}_{10,j}: \alpha_j\neq 0, \beta_j=0;\\
&&H^{}_{01,j}: \alpha_j=0, \beta_j\neq 0.
\end{eqnarray*}
The focus of our study in the context of streaming data is primarily on four testing methods for (\ref{MT-1}),
including Sobel test, adjusted Sobel test, joint significance test and  adjusted joint significance test.

$\bullet$ {\bf{Sobel Test}}  \cite[]{Sobel-1982}.  The Sobel test is widely employed in the mediation analysis literature to evaluate the statistical significance of mediation effects. When the $k$th data batch $\bm{D_k}$ is available, an estimated standard error of $\tilde{\alpha}_{j}^{(k)} \tilde{\beta}_{j}^{(k)}$
has the following expression:
\begin{align}\label{sig_ab}
\tilde{\sigma}_{\alpha_j\beta_j}^{(k)}
& =
{\sqrt{(\tilde{\alpha}_{j}^{(k)})^2 
		\left(\tilde{\sigma}_{\beta_j}^{(k)}\right)^2 + 
		(\tilde{\beta}_{j}^{(k)})^{2} 
		\left(\tilde{\sigma}_{\alpha_j}^{(k)}\right)^2}},~j=1,\ldots,p,
\end{align}
where $\tilde{\beta}_{j}^{(k)}$, ${\tilde{\sigma}_{\beta_j}^{(k)}}$, $\tilde{\alpha}_{j}^{(k)}$ and ${\tilde{\sigma}_{\alpha_j}^{(k)}}$ are given in (\ref{betahat}), (\ref{betaSE}), (\ref{EQ-2-6}) and (\ref{EQ-2-7}), respectively. The Sobel's statistics is defined as 
\begin{align}\label{Tsobel}
T_{Sobel,j}^{(k)}
& =\dfrac{\tilde{\alpha}_{j}^{(k)} \tilde{\beta}_{j}^{(k)} }
{\tilde{\sigma}_{\alpha_j\beta_j}^{(k)}},~j=1,\ldots,p,
\end{align}
where $\tilde{\sigma}_{\alpha_j\beta_j}^{(k)}$ is given in (\ref{sig_ab}). Under $H_{0j}$, the Sobel test assumes that the asymptotic distribution of $T_{Sobel,j}^{(k)}$ is $N(0,1)$. Therefore, the p-value of Sobel test is 
\begin{align*}
P_{Sobel,j}^{(k)} = 2 \{ 1- \Phi_{N(0,1)} (|T_{Sobel,j}^{(k)}|) \},
\end{align*}
where $\Phi_{N(0,1)}(\cdot)$ is the cumulative distribution function of $N(0,1)$. Under the significance level $\delta$, we will reject $H_{0j}$ if $P_{Sobel,j}^{(k)} < {\delta}/{p}$, where the cutoff is Bonferroni corrected for controlling the family-wise error rate (FWER). That is to say, the estimated index set of significant mediators with Sobel test  is
$\hat{\Omega}_{Sobel} =\{j: P_{Sobel,j}^{(k)} < {\delta}/{p}, j=1,\ldots,p \}$. In addition, the Sobel-type $100(1-\delta)\%$  confidence interval for $\alpha_j\beta_j$ is given by
\begin{eqnarray}\label{CI-sobel}
{\rm CI_{Sobel}}=[\tilde{\alpha}_{j}^{(k)} \tilde{\beta}_{j}^{(k)}  - N_{1-\delta/2}(0,1)\tilde{\sigma}_{\alpha_j\beta_j}^{(k)},~\tilde{\alpha}_{j}^{(k)} \tilde{\beta}_{j}^{(k)}  + N_{1-\delta/2}(0,1)\tilde{\sigma}_{\alpha_j\beta_j}^{(k)}],
\end{eqnarray}
where $N_{1-\delta/2}(0,1)$ is the $(1-\delta/2)$-quantile of $N(0,1)$.\\

$\bullet$ {\bf{Adjusted Sobel Test}}  \cite[]{AJS-2024}.  The Sobel test exhibits an excessively conservative type I error, particularly in cases where both $\alpha_j =0$ and $\beta_j =0$. The reason is that the Sobel statistic $T_{Sobel,j}^{(k)}$ follows an asymptotic normal distribution $N(0,1)$ under both $H_{01,j}$ and $H_{10,j}$, whereas it has an asymptotic distribution of $N(0,1/4)$ in the case of $H_{00,j}$. To address this issue, \cite{AJS-2024} proposed a novel adjusted Sobel test by differentiating $H_{00,j}$ from $H_{01,j}$ and $H_{10,j} $ reasonably. For $j=1, \ldots, p$, we denote
\begin{align}\label{Tab-13}
T_{\alpha_{j}}^{(k)}
= 
\dfrac{\tilde{\alpha}_{j}^{(k)}}
{\tilde{\sigma}_{\alpha_{j}}^{(k)}}, ~ T_{\beta_j}^{(k)}
=\dfrac{\tilde{\beta}_{j}^{(k)}}
{\tilde{\sigma}_{\beta_{j}}^{(k)} },
\end{align}
where $\tilde{\beta}_{j}^{(k)}$, ${\tilde{\sigma}_{\beta_j}^{(k)}}$, $\tilde{\alpha}_{j}^{(k)}$ and ${\tilde{\sigma}_{\alpha_j}^{(k)}}$ are given in (\ref{betahat}), (\ref{betaSE}), (\ref{EQ-2-6}) and (\ref{EQ-2-7}), respectively.  From \cite{AJS-2024}, the p-value of adjusted Sobel test is defined as
\begin{align*}
P_{ASobel,j}^{(k)}= \left\{
\begin{aligned}
2 \{ 1- \Phi_{N(0,1)} (|(T_{Sobel,j}^{(k)}|) \},
 \quad \max\{|
T_{\alpha_j}^{(k)}|, 
|T_{\beta_j}^{(k)}
|\} \geq \lambda_{N_k}, 
\\
2 \{ 1- \Phi_{N(0,1/4)} (|T_{Sobel,j}^{(k)}|) \} ,
\quad \max\{|T_{\alpha_j}^{(k)}|, 
|T_{\beta_j}^{(k)}|\} < \lambda_{N_k}, 
\\
\end{aligned}
\right.
\end{align*}
where $T_{Sobel,j}^{(k)}$ is given in (\ref{Tsobel}); 
$\Phi_{N(0,1/4)}(\cdot)$ is the cumulative distribution function of $N(0,1/4)$; $\lambda_{N_k} = {\sqrt{N_k}}/{\log(N_k)}$ and $N_k$ is the total sample size of aggregated streaming data up to batch $k$; $T_{\alpha_{j}}^{(k)}$ and $T_{\beta_j}^{(k)}$ are given in (\ref{Tab-13}). Under the significance level $\delta$, an estimated index set of significant mediators with adjusted Sobel test is $\hat{\Omega}_{ASobel} =\{j: P_{ASobel,j}^{(k)} < {\delta}/{p}, j=1,\ldots,p\}$.

Based on \cite{AJS-2024},  the adjusted Sobel-type $100(1-\delta)\%$ confidence interval for $\alpha_j\beta_j$ is
\begin{eqnarray}\label{CI-Asobel}
&&{\rm CI_{ASobel}}\\
&&=\left\{\begin{array}{l} [\tilde{\alpha}_{j}^{(k)} \tilde{\beta}_{j}^{(k)} - N_{1-\delta/2}(0,1)\tilde{\sigma}_{\alpha_j\beta_j}^{(k)},~\tilde{\alpha}_{j}^{(k)} \tilde{\beta}_{j}^{(k)} + N_{1-\delta/2}(0,1)\tilde{\sigma}_{\alpha_j\beta_j}^{(k)}],~~ \max\{|T_{\alpha_j}^{(k)}|, |T_{\beta_j}^{(k)}|\} \geq \lambda_{N_k},\nonumber\\
{[\tilde{\alpha}_{j}^{(k)} \tilde{\beta}_{j}^{(k)} - N_{1-\delta/2}(0,1/4)\tilde{\sigma}_{\alpha_j\beta_j}^{(k)},~\tilde{\alpha}_{j}^{(k)} \tilde{\beta}_{j}^{(k)} + N_{1-\delta/2}(0,1/4)\tilde{\sigma}_{\alpha_j\beta_j}^{(k)}]},~~\max\{|T_{\alpha_j}^{(k)}|, |T_{\beta_j}^{(k)}|\} < \lambda_{N_k},
\end{array}\right.
\end{eqnarray}
where $N_{1-\delta/2}(0,1/4)$ is the $(1-\delta/2)$-quantile of $N(0,1/4)$.\\

$\bullet$ {\bf{Joint Significance  Test}}  \cite[]{David-PM-2002}. The joint significance (JS) test, also known as the MaxP test, is another widely used method for mediation analysis. It rejects $H_{0j}$ when both ${\alpha_j} = 0$ and ${\beta_j} = 0$ simultaneously. Let
\begin{align*}
P_{\alpha_j}^{(k)} &= 
2 \{ 1- \Phi_{N(0,1)} (|T_{\alpha_j}^{(k)}|)
\},\\
P_{\beta_j}^{(k)}& = 
2 \{ 1- \Phi_{N(0,1)} (| 
T_{\beta_j}^{(k)}
|)\},
\end{align*}
where $T_{\alpha_{j}}^{(k)}$ and $T_{\beta_j}^{(k)}$ are given in (\ref{Tab-13}). By \cite{David-PM-2002}, 
the JS test statistic is defined as
\begin{align}\label{PJS}
P_{JS,j}^{(k)}= \max \{P_{\alpha_j}^{(k)} , P_{\beta_j}^{(k)} \}, ~ j=1,\ldots, p.
\end{align}
To control the FWER, an estimated index set of significant mediators with JS test is 
$\hat{\Omega}_{JS} =\{j:P_{JS,j}^{(k)} < {\delta}/{p}, j=1,\ldots,p\} $, where $\delta$ 
is the significance level. \\

$\bullet$ {\bf{Adjusted Joint Significance  Test}} \cite[]{AJS-2024}.  The JS test suffers from overly conservative type I error,  especially when  both $\alpha_j =0$ and $\beta_j =0$. Under $H_{0j}$, the JS test regards $P_{JS,j}^{(k)}$ as a uniform random variable over $(0,1)$, where
$P_{JS,j}^{(k)}$ is given in (\ref{PJS}). However, the actual distribution of $P_{JS,j}^{(k)}$ is not uniform under the component null hypotheses $H_{00,j}$ (see Figure 3 in \cite {AJS-2024}), which is the reason of  the conservative performance of JS test. To deal with this problem, 
\cite{AJS-2024} proposed an adjusted joint significance (AJS) test with p-value:
\begin{align}
\label{EQ-2-10}
P_{AJS,j}^{(k)}= \left\{
\begin{aligned}
P_{JS,j}^{(k)},\quad \quad \max\{|
T_{\alpha_j}^{(k)}|, 
|T_{\beta_j}^{(k)}
|\} \geq \lambda_{N_k}, 
\\
(P_{JS,j}^{(k)})^2,  \quad \max\{|
T_{\alpha_j}^{(k)}|, 
|T_{\beta_j}^{(k)}
|\} < \lambda_{N_k}, 
\\
\end{aligned}
\right.
\end{align}
where  $T_{\alpha_{j}}^{(k)}$ and $T_{\beta_j}^{(k)}$ are given in (\ref{Tab-13}); $\lambda_{N_k} = {\sqrt{N_k}}/{\log(N_k)}$ and $N_k$ is the total sample size of aggregated streaming data up to batch $k$; $P_{JS,j}^{(k)}$ is defined in (\ref{PJS}). From Theorem 2 of \cite {AJS-2024}, the
AJS  test can effectively control the FWER of multiple mediation testing as presented in (\ref{MT-1}). An estimated index set of significant mediators with AJS test is
$\hat{\Omega}_{AJS} =\{j: P_{AJS,j}^{(k)} <{\delta}/{p}, j=1,\ldots,p \}$, where $P_{AJS,j}^{(k)}$ is defined in \eqref{EQ-2-10}, and $\delta$ is the significance level.

\section{Logistic Mediation Model}
\setcounter{equation}{0}
In this section, we explore the online updating estimation and testing procedures for logistic mediation models with streaming data. 
The counterfactual-based mediation models that involve continuous mediators and a binary outcome are
\begin{eqnarray}
P\{ Y(x,\boldsymbol{m}) = 1\} &=& 
\dfrac{\exp \{ c + \gamma x +
	\beta_{1} m_1 + \cdots + \beta_{p} m_p
	+ \bm{\theta}^{'}  \bm{Z}
	 \}}
{ 1 + \exp \{ c + \gamma x +
	\beta_{1} m_1 + \cdots + \beta_{p} m_p
	+ \bm{\theta}^{'}  \bm{Z}
	  \} },\label{EQ-2-11}\\
M_{j}(x) &=&\alpha_{j} x + \bm{\eta}_{j}^{'} \bm{Z} + e_{j}, j=1,\ldots, p,
\label{EQ-2-12}
\end{eqnarray}
where $Y(x,\boldsymbol{m}) \in \{0,1\} $ is the binary outcome and $c$ is the intercept term; 
other variables and parameters are
similarly defined as the linear mediation models in Section \ref{sec-2}. 
Following \cite{Odds_ratios},   we adopt the
odds ratio scale for definitions of direct and indirect effects, where the natural direct effect on the odds ratio scale ($\rm {NDE^{OR}}$) has the form
\begin{eqnarray*}\label{ORNED}
\rm {NDE^{OR}} = \frac{P\{Y(x,\mathbf{M}(x^*))=1\}/[1-P\{Y(x,\mathbf{M}(x^*))=1\}]}{P\{Y(x^*,\mathbf{M}(x^*))=1\}/[1-P\{Y(x^*,\mathbf{M}(x^*))=1\}]};
\end{eqnarray*}
the natural indirect effect on the odds ratio scale ($\rm {NIE^{OR}}$) is given by
\begin{eqnarray*}\label{ORNIE}
\rm {NIE^{OR}} = \frac{P\{Y(x,\mathbf{M}(x))=1\}/[1-P\{Y(x,\mathbf{M}(x))=1\}]}{P\{Y(x,\mathbf{M}(x^*))=1\}/[1-P\{Y(x,\mathbf{M}(x^*))=1\}]};
\end{eqnarray*}
the total effect on the odds ratio scale ($\rm {TE^{OR}}$) is
\begin{eqnarray*}\label{ORTE}
\rm {TE^{OR}} = \frac{P\{Y(x,\mathbf{M}(x))=1\}/[1-P\{Y(x,\mathbf{M}(x))=1\}]}{P\{Y(x^*,\mathbf{M}(x^*))=1\}/[1-P\{Y(x^*,\mathbf{M}(x^*))=1\}]}.
\end{eqnarray*}
From the above three definitions, we have this expression $\rm {TE^{OR}} = \rm {NIE^{OR}}\cdot\rm {NDE^{OR}}$.  That is  to say, $\log(\rm{TE^{OR}}) = \log(\rm {NIE^{OR}}) + \log(\rm {NDE^{OR}})$ with the log scale. Assuming that conditions (C.1)-(C.4) are satisfied and the event is rare, Theorem 2 of \cite{big-med-2023} indicates that
\begin{eqnarray*}
\rm {NDE^{OR}}&=& \exp\{\gamma(x-x^*)\},\\
\rm {NIE^{OR}}&=& \exp\left\{\sum_{j=1}^p\alpha_j\beta_j(x-x^*)\right\},\\
\rm {TE^{OR}} &=&\exp\left\{\Big(\gamma + \sum_{j=1}^p\alpha_j\beta_j\Big)(x-x^*)\right\}.
\end{eqnarray*}
Therefore, 
the mediation effect along the causal pathway $X \rightarrow M_j \rightarrow Y$ (Figure \ref{fig-1-1}) can be described by the product-of-coefficients $\alpha_{j} \beta_{j}$.

Below, we first consider the online updating estimation for mediation effects with binary outcomes. In the framework of streaming data, we suppose that the data streams from models (\ref{EQ-2-11}) and (\ref{EQ-2-12}) are represented as
 $ \boldsymbol{D_{1}} =\{\boldsymbol{Y}^{(1)}, \boldsymbol{W}^{(1)} \}, 
 \ldots, 
 \boldsymbol{D_{k}} =\{\boldsymbol{Y}^{(k)}, \boldsymbol{W}^{(k)} \}$, 
where $ \boldsymbol{Y}^{(i)}  = \big( Y_{1}^{(i)}, \ldots , Y_{n_i}^{(i)}  \big)^{\bm{'}}$ denotes the binary outcomes from the $i$th batch $\boldsymbol{D}_i$; 
$\boldsymbol{ W}^{(i)} \in \mathbb{R}^{n_i \times (2+p+q)}$ is a matrix consisting of exposures, mediators and covariates, which is denoted as
\begin{align*}
\boldsymbol{W}^{(i)} =\left[ \begin{array}{ccccccccc}
1&X_{{11}}^{(i)} & M_{{11}}^{(i)} & M_{{12}}^{(i)} & \cdots & M_{{1p}}^{(i)} &  Z_{{11}}^{(i)} & \cdots & Z_{{1q}}^{(i)} \\
1&X_{{21}}^{(i)} & M_{{21}}^{(i)} & M_{{22}}^{(i)} & \cdots & M_{{2p}}^{(i)} &  Z_{{21}}^{(i)} & \cdots & Z_{{2q}}^{(i)} \\
\cdots &\cdots & \cdots & \cdots & \cdots & \cdots & \cdots& \cdots & \cdots\\ 
1&X_{{n_i 1}}^{(i)} & M_{{n_i 1}}^{(i)} & M_{{n_i 2}}^{(i)} & \cdots & M_{{n_i p}}^{(i)} &  Z_{{n_i 1}}^{(i)} & \cdots & Z_{{n_i q}}^{(i)} \\
\end{array}\right].
\end{align*}
Let $\boldsymbol{\varpi}_{j}^{(i)}$ as the $j$th row of matrix $\boldsymbol{W^{(i)}}$ and 
$\boldsymbol{ \Gamma} = (c, \gamma,\beta_{1},\ldots,\beta_p,\theta_{1},\ldots,\theta_{q})^{'}$. 
Denote $\boldsymbol{U}^{(i)}$ as the score function with batch $\boldsymbol{D_{i}}$, which is given by
\begin{align*}
\bm{U}^{(i)} 
&=\sum\limits_{j=1}^{n_i} \bm{\varpi_j}^{(i)}
\left\{ Y_{j}^{(i)} - \dfrac{\exp(\bm{\varpi_j}^{(i)} \bm{\Gamma}^{}) }  
{ 1 + \exp(\bm{\varpi_j}^{(i)} \bm{\Gamma^{}}) } 
\right\}.
\end{align*}
The corresponding negative Hessian matrix of  $\boldsymbol{U}^{(i)}$  is 
\begin{align*}
\bm{J}^{(i)}
&= \sum\limits_{j=1}^{n_i} \dfrac{\exp(\bm{\varpi_j}^{(i)} \bm{\Gamma^{}})}
{\{ 1+ \exp(\bm{\varpi_j}^{(i)} \bm{\Gamma^{}}) \}^2} 
\bm{\varpi_j}^{(i)} ( \bm{\varpi_j}^{(i)} )^{'}.
\end{align*}
Let $\boldsymbol{\tilde{J}}^{(k)} = \sum_{i=1}^{k} \boldsymbol{J^{(i)}}$
be the sum of $k$ negative Hessian matrices.  When the $k$th data batch is available, the renewable estimate $\bm{\tilde{\Gamma}}^{(k)}$ can be obtained by solving the incremental estimating equation \cite[]{Lou_2019_Renewable}:
\begin{align*}
\sum\limits_{i =1}^{k-1} \bm{J^{(i)}} (\bm{\tilde{\Gamma}^{(k-1)}} - \bm{\tilde{\Gamma}^{(k)}}) 
+ \bm{U^{(k)}} 
= 0.
\end{align*}
The renewable estimator of parameter $\boldsymbol{\beta}$ is
\begin{align}
\boldsymbol{ \tilde{\beta}}^{(k)}  = (\tilde{\beta}_{1}^{(k)}, \tilde{\beta}_{2}^{(k)}, \ldots , \tilde{\beta}_{p}^{(k)})
= \big( \boldsymbol{ \tilde{\Gamma}}_{3}^{(k)} ,\boldsymbol{ \tilde{\Gamma} }_{4}^{(k)}, \ldots, \boldsymbol{ \tilde{\Gamma}}_{p+2 }^{(k)}   \big),
\end{align}
where  $\boldsymbol{ \tilde{\Gamma}}_{j}^{(k)} $stands for the $j$th item of $\boldsymbol{ \tilde{\Gamma}}^{(k)} $. According to \cite{Lou_2019_Renewable}, the renewable variance matrix of $\boldsymbol{ \tilde{\Gamma}}^{(k)} $ is calculated as 
$$
\bm{\Sigma _{\tilde{\Gamma}^{(k)}}} 
= (\bm{\tilde{J}}^{(k)})^{-1}.
$$
By taking the diagonal of $\boldsymbol{\Sigma_{\tilde{\bm{\Gamma}}^{(k)}}}$,  the estimated standard variance of ${\tilde{\beta }_j}^{(k)}$ is 
\begin{eqnarray}
({\tilde{\sigma}_{\beta_1}^{(k)}},\ldots,{\tilde{\sigma}_{\beta_p}^{(k)}}) =
\big( \sqrt{(\boldsymbol{\Sigma_{\tilde{\bm{\Gamma}}^{(k)}}})_{3,3}}, \ldots, \sqrt{(\boldsymbol{\Sigma_{\tilde{\bm{\Gamma}}^{(k)}}})_{p+2,p+2}}  \big). 
\end{eqnarray}

The renewable estimator for $\boldsymbol{\alpha}$, which characterizes the effects along the pathway $X \rightarrow M_k$, is equivalent to the linear mediation model discussed in Section \ref{sec-2}. Based on the renewable estimators $\tilde{\beta}_{j}^{(k)}$, ${\tilde{\sigma}_{\beta_j}^{(k)}}$, $\tilde{\alpha}_{j}^{(k)}$ and ${\tilde{\sigma}_{\alpha_j}^{(k)}}$, it is straightforward to perform multiple testing:
\begin{align*}
H_{0j}: \alpha_j\beta_j = 0 \leftrightarrow H_{Aj}: \alpha_j\beta_j \neq 0, ~j=1,\ldots,p.
\end{align*}
The Sobel test, adjusted Sobel test, joint significance test, and adjusted joint significance test can be applied in a similar manner as those described in Section \ref{sec22}. The specific details are not provided here. Similarly, the Sobel-type and  adjusted Sobel-type confidence intervals for $\alpha_j\beta_j$'s are given as (\ref{CI-sobel}) and (\ref{CI-Asobel}), respectively.

\section{Simulation Study}
\setcounter{equation}{0}
In this section, we conduct simulations to assess the performance of the proposed online updating method in the settings of linear and logistic mediation models.

\subsection{Parameter Estimation}
First we sequentially generate the data streams $\boldsymbol{D_{1}}, \ldots, \boldsymbol{D_{k}}$ 
from the linear and logistic mediation models, respectively. The sample size ($n_i$) of each data batch $\boldsymbol{D_{i}}$ is uniform across all batches. 
The total sample size of aggregated streaming data up to batch $k$ is $N_k =30000$, where the number of data batch is chosen as $k=10$, 20, 50 and 100, respectively. 
We consider two scenarios for generating the random samples in the settings of linear and logistic mediation models, respectively. 

$\bullet$~{\it Linear Mediation Model with Case 1}: The outcome is generated from 
$Y=  \gamma X +\beta_{1} M_1 + \cdots + \beta_{p} M_p+ \bm{\theta}^{'}  \bm{Z}+ \epsilon$, where
the exposure $X$ follows from $N(0, 2)$, the vector of confounders $\bm{Z}=(Z_{1}, Z_{2})^{\prime}$ with  $Z_1$ and $Z_2$ being independently generated from $N(0, 1)$, the error term $\epsilon$ follows from $N(0, 1)$; $\gamma=0.5$, $\bm{\beta}=(0.15,0.25,0,0,0.15)^{'}$
and $\boldsymbol{\theta}=(0.5,0.5)^{\prime}$. The mediators are generated from 
$M_{j} =  \alpha_{j} X + \bm{\eta}_{j}^{'} \bm{Z} + e_{j}$, $j=1,\ldots, p$, where $\bm{\alpha}=(0.1,0,0,0.35,0.25)^{'}$, $\boldsymbol{\eta_j}=(0.3,0.3)^{\prime}$. The error term
$\mathbf{e} = (e_1,\ldots,e_p)^\prime$ follows from $N(0,\Sigma_e)$ with
$(\Sigma_e)_{i,j} = 0.15^{\vert i - j \vert}$.

$\bullet$~{\it Linear Mediation Model with Case 2}: The exposure $X$ is generated from $B(1,0.5) $
and other settings are the same as Case 1.

$\bullet$~{\it Logistic Mediation Model with Case 3}: The binary outcome $Y\in \{0, 1\}$ is generated from 
\begin{eqnarray*}
P\{ Y = 1\} &=& 
\dfrac{\exp \{\gamma X +
	\beta_{1} M_1 + \cdots + \beta_{p} M_p
	+ \bm{\theta}^{'}  \bm{Z}
	 \}}
{ 1 + \exp \{\gamma X +
	\beta_{1} M_1 + \cdots + \beta_{p} M_p
	+ \bm{\theta}^{'}  \bm{Z}
	  \} },
\end{eqnarray*}
where $\bm{\beta}=(0,0.2,0,0.3,0.25)^{'}$ and $\bm{\alpha}=(0,0.25,0.3,0,0.3)^{'}$. The exposure $X$ follows from $N(0, 2)$, and other settings are the same as Case 1. 

$\bullet$~{\it Logistic Mediation Model with Case 4}:  The exposure $X$ is generated from $B(1,0.5)$
and other settings are the same as Case 3.

The "Full Data Method" involves utilizing the aggregated streaming data up to batch $k$ directly for conducting mediation analysis, as a means of comparison. 
In  Tables \ref{table 1}-\ref{table 4}, we report the following mediation analysis results, including the bias (Bias) given by the difference of sample means of estimate
and the true value, the sampling standard errors (SSE), the average of 
estimated standard errors (ASE), and the 95\% empirical coverage probabilities of Sobel method (${\rm CP_{Sobel}}$) and adjusted Sobel method (${\rm CP_{ASobel}}$). The
Sobel-type and  adjusted Sobel-type confidence intervals for $\alpha_j\beta_j$'s are given in 
(\ref{CI-sobel}) and (\ref{CI-Asobel}), respectively.  All the results are based on 500 repetitions.

It can be seen from Tables \ref{table 1}-\ref{table 4} that the estimators are unbiased and the 
variance estimation given in (\ref{sig_ab}) seems to work well. The coverage probabilities of adjusted Sobel-type confidence intervals are reasonable
and consistent with the normal levels. However, the coverage probability of the $95\%$ Sobel-type confidence interval approaches unity when both $\alpha$ and $\beta$ are equal to zero.
We assessed the computational efficiency of our 
method. The
computations were carried out using R  on a desktop computer with
32GB memory. We restricted the calculations to access one CPU core and recorded
the average CPU time (in seconds) from 500 repetitions. The renewable method consistently demonstrates comparable estimation performance to the full data method across various batch sizes, while offering significantly improved computational efficiency.

\begin{table}[htp]
	\begin{center} 
	\caption{The results of linear mediation model with Case 1. }
	\label{table 1}
	\begin{threeparttable}
\resizebox{1.0\height}{!}{
		\begin{tabular}{*{7}{c}}
		\toprule
\multirow{2}{*}{} 
& & \multicolumn{4}{c}{Renewable Method} 
& \multicolumn{1}{c}{Full Data Method}
\\
\cmidrule(r){3-6} 
 & & $k=10$ & $k=20$    &  $k=50$  &   $k=100$ &  $N_k=30000$\\
\midrule
 & ${\alpha}_1 {\beta}_1$ 
 & 275.44
 & 275.44     
 & 275.44  
 & 275.44  
 & 275.44  \\
 & ${\alpha}_2 {\beta}_2$ 
 & -132.00 
 & -132.00
 & -132.00 
 & -132.00 
 & -132.00
 \\
  Bias $\times 10^{7}$ & ${\alpha}_3 {\beta}_3$ 
  & -5.18   
  & -5.18 
  & -5.18   
  & -5.18  
  & -5.18  
 \\
 & ${\alpha}_4 {\beta}_4$  
 & -319.36
 & -319.36
 & -319.36
 & -319.36
 & -319.36
 \\
 & ${\alpha}_5{\beta}_5$ 
 & -694.11
 &-694.11
 & -694.11
 & -694.11
 & -694.11
 \\
 \midrule
 &${\alpha}_1{\beta}_1$ 
 & 70.62 
 & 70.62   
 & 70.62 
 & 70.62 
 & 3.16 
 \\
 & ${\alpha}_2 {\beta}_2$ 
 & 71.85
 & 71.85
 & 71.85
 & 71.85
 & 3.21
 \\
SSE $\times 10^{5}$ 
& ${\alpha}_3{\beta}_3$ 
& 1.64 
& 1.64
& 1.64
& 1.64
& 0.07
 \\
 & ${\alpha}_4 {\beta}_4$ 
 & 198.62
 & 198.62
 & 198.62
 & 198.62
 & 8.88
 \\
 & ${\alpha}_5 {\beta}_5$ 
 & 159.14
 & 159.14
 & 159.14
 & 159.14
 & 7.12
 \\
  \midrule
  & ${\alpha}_1 {\beta}_1$ 
 & 72.78 
 & 72.78 
 & 72.78 
 & 72.78 
 & 3.25 
 \\
 & ${\alpha}_2 {\beta}_2$   
 & 72.21 
 & 72.21
 & 72.21 
 & 72.21
 & 3.23 
  \\
 ASE $\times 10^{5}$
 & ${\alpha}_3 {\beta}_3$ 
 & 2.14  
 & 2.14  
 & 2.14 
 & 2.14  
 & 0.09\\
 & ${\alpha}_4 {\beta}_4$ 
 & 206.74 
 & 206.74
 & 206.74
 & 206.74
 & 9.25
 \\
 & ${\alpha}_5 {\beta}_5$ 
 & 152.32  
 & 152.32 
 & 152.32 
 & 152.32
 & 6.81\\
  \midrule
   & ${\alpha}_1 {\beta}_1$ 
   & 0.968 & 0.968 & 0.968 & 0.968 & 0.968 \\
   & ${\alpha}_2 {\beta}_2$ & 0.944 & 0.944 & 0.944  & 0.944 & 0.944\\
 ${\rm CP_{Sobel}}$ 
 & ${\alpha}_3 {\beta}_3$ &1  &1 & 1 & 1 & 1 \\
 & ${\alpha}_4 {\beta}_4$ & 0.958 & 0.958  & 0.958  & 0.958  & 0.958  \\
 & ${\alpha}_5 {\beta}_5$ & 0.940 & 0.940 &  0.940 & 0.940 &  0.940\\
   \midrule
  & ${\alpha}_1 {\beta}_1$& 0.968 & 0.968 & 0.968 & 0.968 & 0.968 \\
  & ${\alpha}_2 {\beta}_2$ & 0.944 & 0.944 & 0.944  & 0.944 & 0.944\\
 ${\rm CP_{ASobel}}$ 
 & ${\alpha}_3 {\beta}_3$ & 0.950  &0.950 & 0.950 & 0.950 & 0.950 \\
 & ${\alpha}_4 {\beta}_4$ & 0.958 & 0.958  & 0.958  & 0.958  & 0.958\\
 & ${\alpha}_5 {\beta}_5 $ & 0.940 & 0.940 &  0.940 & 0.940 &  0.940\\
    \midrule
 CPU time & & 0.0211 & 0.0257 &0.0427 & 0.0793 & 0.0802 \\

\bottomrule      

\end{tabular} }

\end{threeparttable}
\end{center}
\end{table}

\begin{table}
	\begin{center} 
	\caption{The results of linear mediation model with Case 2.}
	\label{table 2}
	\begin{threeparttable}
		\resizebox{1.0\height}{!}{
			\begin{tabular}{*{7}{c}}
				\toprule
				\multirow{2}{*}{} &
				& \multicolumn{4}{c}{Renewable Method} 
				& \multicolumn{1}{c}{Full Data Method}
				\\
				\cmidrule(r){3-6} 
				& & $ k=10$ & $k=20$    &  $k=50$  &   $k=100$ &  $N_k=30000$\\
				\midrule
				& ${\alpha}_1 {\beta}_1$ 
				& 75.36 
				& 75.36   
				& 75.36 
				& 75.36 
				& 75.36
				 \\
				& ${\alpha}_2 {\beta}_2$ 
				&-106.95  
				& -106.95 
				& -106.95 
				& -106.95
				& -106.95
				 \\
			Bias$\times 10^{6}$	
			& ${\alpha}_3{\beta}_3$ 
				& 1.91
				& 1.91
				& 1.91
				& 1.91
				& 1.91
				\\
			& ${\alpha}_4 {\beta}_4$ 
				& -14.59
				& -14.59
				& -14.59
				& -14.59
				& -14.59
				\\
				& ${\alpha}_5 {\beta}_5$ 
				& -95.93
			    & -95.93
				&-95.93
				& -95.93
				& -95.93
				\\
				\midrule
		    	& ${\alpha}_1 {\beta}_1$ 
				& 180.75 
				& 180.75
				& 180.75
				& 180.75
				& 8.08
				\\
				& ${\alpha}_2{\beta}_2$ 
				& 271.65 
				& 271.65
				& 271.65
				& 271.65
				& 12.15
				\\
			SSE $\times 10^{5}$	& ${\alpha}_3 {\beta}_3$ 
				& 7.12
				& 7.12
				& 7.12
				& 7.12
				& 0.32
				\\
				& ${\alpha}_4 {\beta}_4$ 
				& 192.57
				& 192.57
				& 192.57
				& 192.57
				& 8.61
				\\
				& ${\alpha}_5 {\beta}_5$ 
				& 221.78
				& 221.78 
				& 221.78
				& 221.78
				& 9.92
				\\
				\midrule
				& ${\alpha}_1 {\beta}_1$  
				& 183.22
				& 183.22
				& 183.22
				& 183.22
				& 8.19
				\\
				& ${\alpha}_2 {\beta}_2$ 
				& 288.52
				& 288.52
				& 288.52
				& 288.52
				& 12.91
				\\
			ASE $\times 10^{5}$	& ${\alpha}_3 {\beta}_3$ 
				& 8.43
				& 8.43
				& 8.43
				& 8.43
				& 0.38
				\\
				& ${\alpha}_4 {\beta}_4$ 
				& 206.74 
				& 206.74 
				& 206.74 
				& 206.74 
				& 9.25
				\\
				& ${\alpha}_5 {\beta}_5$ 
						& 226.27 
			& 226.27 
			& 226.27
			& 226.27
			& 10.12
				\\
				\midrule
			& ${\alpha}_1 {\beta}_1$ 
				& 0.958 & 0.958 & 0.958 & 0.958 & 0.958 \\
				& ${\alpha}_2 {\beta}_2$ 
				& 0.970 & 0.970 & 0.970  & 0.970 & 0.970\\
			${\rm CP_{Sobel}}$ 	& ${\alpha}_3 {\beta}_3$ 
				&1  &1 & 1 & 1 & 1 \\
				& ${\alpha}_4 {\beta}_4$ 
				& 0.966 & 0.966 & 0.966  & 0.966  & 0.966  \\
				& ${\alpha}_5 {\beta}_5$ 
				& 0.948 & 0.948 &  0.948 & 0.948 &  0.948\\
				\midrule
				& ${\alpha}_1 {\beta}_1$ 	& 0.958 & 0.958 & 0.958 & 0.958 & 0.958 
				\\
				& ${\alpha}_2 {\beta}_2$ & 0.970 & 0.970 & 0.970  & 0.970 & 0.970\\
			${\rm CP_{ASobel}}$ 	& ${\alpha}_3 {\beta}_3$ & 0.952 & 0.952 & 0.952  & 0.952 & 0.952\\
				& ${\alpha}_4 {\beta}_4$ & 0.966 & 0.966 & 0.966  & 0.966  & 0.966  \\
				& ${\alpha}_5 {\beta}_5$ 
				& 0.948 & 0.948 &  0.948 & 0.948 &  0.948\\
				\midrule
			CPU time & & 0.0205 & 0.0257 &0.0415 & 0.0665 & 0.0920 \\				
				\bottomrule      
				
		\end{tabular} }
	\end{threeparttable}
\end{center}
\end{table}

\begin{table}
	\begin{center} 
	\caption{The results of logistic mediation model with Case 3. }
	\label{table 3}
	\begin{threeparttable}
		\resizebox{1\height}{!}{
			\begin{tabular}{*{7}{c}}
				\toprule
				\multirow{2}{*}{}&
				& \multicolumn{4}{c}{Renewable Method} 
				& \multicolumn{1}{c}{Full Data Method}
				\\
				\cmidrule(r){3-6} 
			&	& $ k=10$ & $k=20$    &  $k=50$  &   $k=100$ &  $N_k=30000$\\
				\midrule
				& ${\alpha}_1 {\beta}_1$ 
				& 9.37 
				& 9.32 
				& 9.44 
				& 9.42 
				& 9.40 \\
				
				& ${\alpha}_2 {\beta}_2$
				& -1176.25 
				& -1324.10  
				& -1486.76 
				& -1617.85  
				&  -768.09 \\
				
			Bias$\times 10^{7}$	& ${\alpha}_3 {\beta}_3$ 
				& -4173.34 
				& -4152.52 
				& -4133.17 
				& -4112.40  
				& -4190.28 \\
				
			& ${\alpha}_4 {\beta}_4$ 
				& 289.14 
				& 288.95 
				& 288.90 
				& 288.92 
				& 290.47  \\
				& ${\alpha}_5 {\beta}_5$ 
				& -1022.63 
				& -1195.56
				& -1428.74 
				& -1598.39  
				& -369.49 \\
				\midrule
				& ${\alpha}_1 {\beta}_1$ 
				& 3.99 
				& 3.98
				& 3.98 
				& 3.98 
				& 3.99 \\
				
				& ${\alpha}_2 {\beta}_2$ 
				& 355.77 
				& 355.77 
				& 355.64 
				& 355.61 
				& 356.04 \\
				
		SSE$\times 10^{5}$		& ${\alpha}_3 {\beta}_3$ 
				& 450.10
				& 450.08 
				& 450.08 
				& 450.01 
				& 450.37 \\
				
				& ${\alpha}_4 {\beta}_4$ 
				& 91.077 
				& 91.055 
				& 91.028 
				& 91.001 
				& 91.130 \\
				
				& ${\alpha}_5 {\beta}_5$ 
				& 431.10 
				& 430.86 
				& 430.73 
				& 430.65 
				& 431.37 \\
				\midrule
				
				& ${\alpha}_1 {\beta}_1$ 
				& 5.07
				& 5.07
				& 5.07
				& 5.07
				& 5.07\\
				
				& ${\alpha}_2 {\beta}_2$ 
				& 368.97 
				& 368.94 
				& 368.90 
				& 368.84 
				& 368.82 \\
				
		ASE $\times 10^{5}$	 	& ${\alpha}_3 {\beta}_3$ 
				& 433.74 
				& 433.71 
				& 433.66 
				& 433.60 
				& 433.59 \\
	
				 & ${\alpha}_4 {\beta}_4$
				& 86.70
				& 86.68
				& 86.65
				& 86.63
				& 86.78\\
				
				& ${\alpha}_5 {\beta}_54$
				& 440.37 
				& 440.33 
				& 440.27 
				& 440.20 
				& 440.22 \\
				\midrule
				 & ${\alpha}_1 {\beta}_1$	&1  &1 & 1 & 1 & 1\\
				& ${\alpha}_2 {\beta}_2$& 0.952 & 0.952 & 0.952  & 0.952 & 0.952\\
				${\rm CP_{Sobel}}$ & ${\alpha}_3 {\beta}_3$& 0.938  &0.938 & 0.938 & 0.938 & 0.938 \\
				& ${\alpha}_4 {\beta}_4$& 0.936 & 0.936 & 0.936 & 0.936 & 0.936 \\
				& ${\alpha}_5{\beta}_5$& 0.962 & 0.962 &  0.962 & 0.962 &  0.962\\
				\midrule
				
				& ${\alpha}_1 {\beta}_1$& 0.958 & 0.958 & 0.958 & 0.958 & 0.958 \\
				& ${\alpha}_2 {\beta}_2$& 0.952 & 0.952 & 0.952  & 0.952 & 0.952\\
			${\rm CP_{ASobel}}$& ${\alpha}_3 {\beta}_3$& 0.938  &0.938 & 0.938 & 0.938 & 0.938 \\
				& ${\alpha}_4 {\beta}_4$& 0.936 & 0.936 & 0.936 & 0.936 & 0.936 \\
				& ${\alpha}_5 {\beta}_5$& 0.962 & 0.962 &  0.962 & 0.962 &  0.962\\
				\midrule
				
				CPU time & & 0.0554 & 0.0671 &0.1013 & 0.1486 & 0.2375 \\
				
				\bottomrule      
				
		\end{tabular} }
		
	\end{threeparttable}
\end{center}
\end{table}

\begin{table}
	\begin{center} 
	\caption{The results of logistic mediation model with Case 4. }
	\label{table 4}
	\begin{threeparttable}
	\resizebox{1\height}{!}{
		\begin{tabular}{*{7}{c}}
			\toprule
			\multirow{2}{*}{}&
			& \multicolumn{4}{c}{Renewable Method} 
			& \multicolumn{1}{c}{Full Data Method}
			\\
			\cmidrule(r){3-6} 
			&	& $ k=10$ & $k=20$    &  $k=50$  &   $k=100$ &  $N_k=30000$\\
			\midrule
				 & ${\alpha}_1 {\beta}_1$
				& 2.80 
				& 2.77  
				& 2.75 
				& 2.72  
				& 2.05 \\
				
				 & ${\alpha}_2 {\beta}_2$
				& -134.81 
				& -148.01  
				& -164.95 
				& -176.35  
				& -67.12 \\
				
			Bias$\times 10^{6}$	 & ${\alpha}_3 {\beta}_3$
				& 176.55 
				& 176.29 
				& 177.48 
				& 177.36 
				& 170.62 \\
				 & ${\alpha}_4 {\beta}_4$
				& 31.69
				& 31.64
				& 31.59
				& 31.55
				& 33.78\\
				 & ${\alpha}_5 {\beta}_5$
				& -260.83
				& -278.49 
				& -303.72 
				& -319.89 
				& -216.65\\
				\midrule
				& ${\alpha}_1 {\beta}_1$
				& 1.02
				& 1.02
				& 1.02
				& 1.33
				& 1.43\\
				 & ${\alpha}_2 {\beta}_2$
				& 371.87
				& 371.66 
				& 371.65 
				& 375.04 
				& 399.11 \\
				SSE  $\times 10^{4}$& ${\alpha}_3 {\beta}_3$
				& 404.27 
				& 404.22
				& 404.14 
				& 401.97 
				& 405.52 \\
				
				& ${\alpha}_4 {\beta}_4$
				& 251.99
				& 251.93 
				& 251.85
				& 245.42
				& 346.01\\
				
				& ${\alpha}_5 {\beta}_5$
				& 459.52 
				& 459.31 
				& 459.20 
				& 451.74 
				& 500.39 \\
				
				\midrule
				& ${\alpha}_1 {\beta}_1$
				& 1.33 
				& 1.33 
				& 1.33
				& 1.33 
				& 1.94 \\
				
				& ${\alpha}_2 {\beta}_2$
				& 37.51 
				& 37.51
				& 37.51
				& 37.51
				& 40.95\\
				
			ASE $\times 10^{4}$	& ${\alpha}_3 {\beta}_3$
				& 40.20
				& 40.20
				& 40.20
				& 40.20
				& 40.25\\
				
				& ${\alpha}_4 {\beta}_4$
				& 24.56
				& 24.56
				& 24.55
				& 24.54
				& 34.76\\
				
				 & ${\alpha}_5{\beta}_5$
				& 45.19
				& 45.19
				& 45.18
				& 45.17
				& 49.57\\
				
				\midrule
				  & ${\alpha}_1 {\beta}_1$	&1  &1 & 1 & 1 & 1\\
				  
				 &  ${\alpha}_2 {\beta}_2$& 0.950 & 0.950 & 0.950  & 0.952 & 0.964\\
				 
				${\rm CP_{Sobel}}$  & ${\alpha}_3 {\beta}_3$
				& 0.942  &0.944 & 0.944 & 0.946 & 0.942 \\
				 
				& ${\alpha}_4 {\beta}_4$
				& 0.938 & 0.938 & 0.938 & 0.938 & 0.942 \\
				 
				 & ${\alpha}_5 {\beta}_5$
				 & 0.946 & 0.946 &  0.944 & 0.946 &  0.946\\
				\midrule
				
			  & ${\alpha}_1 {\beta}_1$& 0.954 & 0.954 & 0.954 & 0.956 & 0.958 \\
			  
				 & ${\alpha}_1 {\beta}_2$& 0.950 & 0.950 & 0.950  & 0.952 & 0.954\\
				 
		${\rm CP_{ASobel}}$	& ${\alpha}_3 {\beta}_3$	& 0.942  &0.944 & 0.944 & 0.946 & 0.942 \\
			
				 & $\hat{\alpha}_4 \hat{\beta}_4$& 0.938 & 0.938 & 0.938 & 0.938 & 0.942 \\
				 
			 & ${\alpha}_5 {\beta}_5$	& 0.946 & 0.946 &  0.944 & 0.946 &  0.946\\
				\midrule
				
				CPU time & & 0.0533 & 0.0635 & 0.0986 & 0.1545 & 0.2463 \\
				
				\bottomrule      
				
		\end{tabular} }
		
	\end{threeparttable}
\end{center}
\end{table}

\subsection{Mediation Testing}
The performance of our online updating method for multiple mediation test is evaluated in this section. We generate the data streams $\boldsymbol{D_{1}}, \ldots, \boldsymbol{D_{k}}$ 
from the linear and logistic mediation models. The sample size of each data batch $\boldsymbol{D_{i}}$ is equal across all batches. 
The total sample size of aggregated streaming data up to batch $k$ is $N_k =5000$, where the number of data batch is chosen as $k=5$, 10 and 15, respectively.


$\bullet$~{\it Linear Mediation Model with Case 5}: The outcome is generated from 
$Y=  \gamma X +\beta_{1} M_1 + \cdots + \beta_{p} M_p+ \bm{\theta}^{'}  \bm{Z}+ \epsilon$, where
the exposure $X$ follows from $B(1, 0.5)$, the vector of confounders $\bm{Z}=(Z_{1}, Z_{2})^{\prime}$ with  $Z_1$ and $Z_2$ being independently generated from $N(0, 1)$, the error term $\epsilon$ follows from $N(0, 1)$; $\gamma=0.5$, $\boldsymbol{\theta}=(0.5,0.5)^{\prime}$, $\beta_{1}= 0.15,  
\beta_{2}=0.15,    
\beta_{3}=0.08,    
\beta_{4}=0,      
\beta_{5}=0.35 $
and   $\beta_{i}=0 $ for $i$ = 6, 7, 8, 9, 10. 
The mediators are generated from 
$M_{j} =  \alpha_{j} X + \bm{\eta}_{j}^{'} \bm{Z} + e_{j}$, $j=1,\ldots, 10$, where $\boldsymbol{\eta_j}=(0.3,0.3)^{\prime}$, $\alpha_{1}=0.1,
\alpha_{2}=0.1,
\alpha_{3}=0.1,
\alpha_{4}=0.3,
$ and   $\alpha_{i}=0 $ for $i$ = 5, 6, 7, 8, 9, 10. The error term
$\mathbf{e} = (e_1,\ldots,e_p)^\prime$ follows from $N(0,\Sigma_e)$ with
$(\Sigma_e)_{i,j} = 0.15^{\vert i - j \vert}$.

\begin{table}
	\begin{center} 
	\caption{The results of linear mediation model with Case 5. }
	\label{table 5}
	\begin{threeparttable}
			\begin{tabular}{*{6}{c}}
				\toprule
				\multirow{2}{*}{}
					&\multicolumn{1}{c} {}
				&\multicolumn{3}{c}{Renewable Method}
				& \multicolumn{1}{c}{Full Data Method}
				\\
				\cmidrule(r){3-5} 
				&& $k=5$ & $k=10$    &  $k=15$  &   $N_k=5000$ \\
				\midrule
				FWER & Sobel & 0.010 & 0.010 & 0.010 & 0.010\\
				& ASobel & 0.028 & 0.028 & 0.028 & 0.028\\
				& JS & 0.014 & 0.014 & 0.014 & 0.014\\
				& AJS & 0.032 & 0.032 & 0.032 & 0.032\\
				\midrule
				Power & Sobel & 0.7080 & 0.7080 & 0.7080 & 0.7080 \\
				& ASobel & 0.7993 & 0.7993 & 0.7993 & 0.7993\\
				& JS & 0.7567 & 0.7567  & 0.7567  & 0.7567 \\
				& AJS & 0.8160 & 0.8160 & 0.8160 & 0.8160\\
				\midrule
				CPU time & & 0.0110 & 0.0155 & 0.0190 & 0.0221 \\
				\bottomrule      
				
		\end{tabular} 
		
	\end{threeparttable}
\end{center}
\end{table}

$\bullet$~{\it Linear Mediation Model with Case 6}: The exposure $X$ is generated from $N(0,2)$. The regression coefficients are chosen as $\alpha_{1}=0.06,
\alpha_{2}=0.055,
 \alpha_{3}=0.06,
 \alpha_{4}=0.3
 $ and   $\alpha_{i}=0 $ for $i$ = 5, 6, 7, 8, 9, 10; 
 $\beta_{1}= 0.05,  
 \beta_{2}=0.06,     
 \beta_{3}=0.05,    
 \beta_{4}=0,      
 \beta_{5}=0.25 $
 and   $\beta_{i}=0 $ for $i$ = 6, 7, 8, 9, 10. Other settings are the same as those of Case 5.

\begin{table}
\begin{center} 

	\caption{The results of linear mediation model with Case 6. }
	\label{table 6}
	\begin{threeparttable}
	
			\begin{tabular}{*{6}{c}}
				\toprule
				\multirow{2}{*}{}
					&\multicolumn{1}{c} {}
				&\multicolumn{3}{c}{Renewable Method}
				& \multicolumn{1}{c}{Full Data Method}
				\\
				\cmidrule(r){3-5} 
			&	& $ k=5$ & $k=10$    &  $k=15$  &   $N_k=5000$ \\
				\midrule
FWER & Sobel & 0.006 & 0.006 & 0.006 & 0.006\\
     & ASobel &0.034 & 0.034 & 0.034 & 0.034\\
     & JS & 0.008 & 0.008 & 0.008 & 0.008\\
     & AJS & 0.036 & 0.036 & 0.036 & 0.036\\
     \midrule
 Power & Sobel & 0.7647 & 0.7647 & 0.7647 & 0.7647 \\
       & ASobel & 0.8673 & 0.8673 & 0.8673 & 0.8673\\
       & JS & 0.8067 & 0.8067  & 0.8067  & 0.8067 \\
       & AJS & 0.8780 & 0.8780 & 0.8780 & 0.8780\\
      \midrule
CPU time & & 0.0101 & 	0.0147 & 0.0195 &	0.0219 \\
\bottomrule      

\end{tabular} 

\end{threeparttable}

\end{center}
\end{table}


$\bullet$~{\it Logistic Mediation Model with Case 7}: The binary outcome $Y\in \{0, 1\}$ is generated from 
\begin{eqnarray*}
P\{ Y = 1\} &=& 
\dfrac{\exp \{\gamma X +
	\beta_{1} M_1 + \cdots + \beta_{p} M_p
	+ \bm{\theta}^{'}  \bm{Z}
	 \}}
{ 1 + \exp \{\gamma X +
	\beta_{1} M_1 + \cdots + \beta_{p} M_p
	+ \bm{\theta}^{'}  \bm{Z}
	  \} },
\end{eqnarray*}
where $\beta_{1}= 0.125,  
\beta_{2}=0.1,     
\beta_{3}=0.1,    
\beta_{4}=0.4   $
and   $\beta_{i}=0 $ for i = 5, 6, 7, 8, 9, 10;
 $\alpha_{1}=0.2,
\alpha_{2}=0.25,
\alpha_{3}=0.25,
\alpha_{5}=0.3,
$ and   $\alpha_{i}=0 $ for i = 4, 6, 7, 8, 9, 10.   The exposure $X$ is generated from $B(1, 0.5)$, and other settings are the same as Case 5.

\begin{table}
	\begin{center}		

	\caption{The results of logistic mediation model with Case 7.}
	\label{table 7}
	\begin{threeparttable}
		
			\begin{tabular}{*{6}{c}}
				\toprule
				\multirow{2}{*}{}
				& \multicolumn{1}{c} {}
				&\multicolumn{3}{c}{Renewable Method} 
				& \multicolumn{1}{c}{Full Data Method}
				\\
				\cmidrule(r){3-5} 
				& 	& $ k=5$ & $k=10$    &  $k=15$  &   $N_k=5000$ \\
				\midrule
FWER & Sobel & 0.012 & 0.012 & 0.012 & 0.012\\
	& ASobel & 0.034 & 0.034 & 0.034 & 0.034\\
	& JS & 0.014     & 0.014 & 0.014 & 0.014\\
	& AJS & 0.042    & 0.042 & 0.042 & 0.042\\
			\midrule
			
Power & Sobel & 0.6213 & 0.6207 & 0.6193 & 0.6280 \\
	& ASobel & 0.7633 & 0.7620 & 0.7513 & 0.7680\\
	& JS & 0.6847 & 0.6840  & 0.6847  & 0.6927 \\
	& AJS & 0.7780 & 0.7767 & 0.7773 & 0.7840\\
	
	\midrule
CPU time & & 0.0132 & 0.0181 & 0.0209 & 0.0502\\
\bottomrule      

\end{tabular} 

\end{threeparttable}
	\end{center}
\end{table}

$\bullet$~{\it Logistic Mediation Model with  Case 8}: The exposure $X$ is generated from $N(0,2)$, and we set 
the  regression coefficients as $\alpha_{1}=0.055,
\alpha_{2}=0.06,
\alpha_{3}=0.07,
\alpha_{5}=0.3,
$ and   $\alpha_{i}=0 $ for i = 4, 6, 7, 8, 9, 10;
$\beta_{1}= 0.125,  
\beta_{2}=0.115,      
\beta_{3}=0.105,     
\beta_{4}=0.4   $
and   $\beta_{i}=0 $ for i = 5, 6, 7, 8, 9, 10.  Other settings are the same as those of Case 7.

\begin{table}
	\begin{center}
	\caption{The results of logistic mediation model with Case 8. }
	\label{table 8}
	\begin{threeparttable}
		
			\begin{tabular}{*{6}{c}}
				\toprule
				\multirow{2}{*}{}
				&\multicolumn{1}{c} {}
				&\multicolumn{3}{c}{Renewable Method}
				& \multicolumn{1}{c}{Full Data Method}
				\\
				\cmidrule(r){3-5} 
				&	& $ k=5$ & $k=10$    &  $k=15$  &   $N_k=5000$ \\
				\midrule
				FWER & Sobel & 0.008 & 0.008 & 0.008 & 0.008\\
				& ASobel & 0.036 & 0.034 & 0.034 & 0.036\\
				& JS & 0.010     & 0.010 & 0.010 & 0.010\\
				& AJS & 0.040   & 0.040 & 0.040 & 0.042\\
				\midrule
				
				Power & Sobel & 0.6480 & 0.6467 & 0.6453 & 0.6553 \\
				& ASobel & 0.7640 & 0.7647 & 0.7633 & 0.7687\\
				& JS & 0.7093 &  0.7020 & 0.7040  & 0.7107 \\
				& AJS &  0.7827 & 0.7787& 0.7807 & 0.7840\\
				\midrule
				CPU time & & 0.0130 & 0.0167 & 0.0215 &	0.0354\\
				\bottomrule      
				
		\end{tabular} 
		
	\end{threeparttable}
\end{center}
\end{table}

In the framework of streaming data, we are interested in the multiple mediation testing problem:
\begin{align*}
H_{0j}: \alpha_j\beta_j = 0 \leftrightarrow H_{Aj}: \alpha_j\beta_j \neq 0, ~j=1,\ldots,10,
\end{align*}
where the significance level is $\delta = 0.05$. The Sobel test, adjusted Sobel test (ASobel), joint significance test (JS), and adjusted joint significance test (AJS) are considered in this study, as discussed in Section \ref{sec22}. For comparison, we  also use the aggregated streaming data up to batch $k$ directly for conducting mediation analysis (denoted as "Full Data Method").
All the results are based on 500 repetitions.

The FWERs and Powers of the Sobel, ASobel, JS, and AJS methods are reported in Tables $\ref{table 5}$-$\ref{table 8}$ for mediation tests conducted using both the renewable method and full data method. The results presented in Tables $\ref{table 5}$-$\ref{table 8}$ demonstrate that ASobel outperforms Sobel in terms of both FWERs and Powers, while AJS method shows superior performance compared to JS method. We also evaluate the computational efficiency of our approach. The computations are performed using R on a desktop computer with 32GB memory. We restrict the calculations to utilize only one CPU core and record the average CPU time (in seconds) from 500 repetitions. The renewable method consistently exhibits comparable FWERs and Powers as the full data method across various batch sizes, while offering significantly enhanced computational efficiency.

\section{Real Data Example}
\subsection{Continuous Outcomes}
The application of our method to a substantial dataset on the P2P lending platform is presented in this section.
The acronym P2P stands for peer to peer, representing a novel form of network lending platform. 
 The dataset is about the loan transaction data of Lending Club Company from 2007 to 2015, which is publicly available at \url{ https://www.kaggle.com/wendykan/lending-club-data}. \cite{herzenstein2008-SMR}  revealed a significant association between the borrower's housing property and interest rates.
The total loan amount and repayment period are expected to have a reasonable correlation with the borrower's housing property, while the interest rate is anticipated to be correlated with both the loan amount and repayment period. Therefore, we are committed to investigating whether the loan amount and repayment period play a significant role in establishing a causal relationship between the borrower's housing property and the interest rate.
The resulting total sample size, denoted as $N_k$, was obtained after the removal of missing values from the dataset and amounted to 842,322.

The exposure variable $X$ in our analysis represents the borrower's housing property, encompassing rental properties, owned properties, and mortgaged properties. Specifically, we assign a value of $-1$ to rented or mortgaged properties and a value of $1$ to owned properties.
The outcome variable $Y$ represents the interest rate on the loan. It is anticipated that the categories "rent" or "mortgage" would exhibit an inverse impact on $Y$ compared to the category "own". In order to characterize this attribute, we represent the binary exposure variable $X$ as -1 and 1 instead of encoding it as 0 and 1.

The two mediators under consideration are as follows: Mediator $M_{1}$ represents the total loan amount, measured in units of ten thousand dollars. On the other hand, Mediator $M_2$ signifies the duration of loan repayment, with a value of 1 for a 36-month period and another value for a 60-month period. 
 Furthermore, we define the confounding variables $Z_1$ as the borrower's annual income (measured in units of ten thousand dollars) and $Z_2$ as the borrower's years of work experience. Our objective is to investigate whether there are significant mediated effects of $M_1$ and $M_2$ on the pathways connecting housing status ($X$) with loan interest rates ($Y$).

The linear mediation models are initially applied to the full dataset, which encompasses the subsequent expressions:
\begin{align*}
Y &= 29.667 - 0.025 X + 0.019 M_1 + 0.066 M_2 - 0.009 Z_1 - 3.169 Z_2 + \epsilon ,
\\
M_1 &= 1.745 - 0.009 X + 0.043 Z_1 - 0.114 Z_2 + e_1 ,
\\
M_2 &= 2.082 - 0.009 X + 0.006 Z_1 - 0.159 Z_2 + e_2.
\end{align*}
The full data is divided into $k$ blocks, denoted as $\boldsymbol{D_{1}}, \ldots, \boldsymbol{D_{k}}$, with each block having a sample size of $n_k=N_k/k$. We employ our proposed online updating methods to perform mediation analysis within the framework of streaming data. The value of $k$ is chosen as 10, 50, 100, and 500 respectively.

In Table \ref{table 9}, we report the estimates, standard
errors, p-values and 95\% confidence intervals of mediation effects using the renewable method and the full data method. The provided CPU time (in seconds) is also utilized for evaluating the computational efficiency, utilizing the same computing device as used in the simulation section. The findings presented in Table \ref{table 9} suggest that both the loan amount ($M_1$) and repayment period ($M_2$) can be regarded as two significant mediators in the pathways linking housing status ($X$) to loan interest rates ($Y$). The renewable approach consistently demonstrates comparable estimation to the full data method across various batch sizes, while providing significantly enhanced computational efficiency. The confidence interval of the ASobel method is significantly shorter compared to that of the Sobel method, indicating a higher level of efficiency in terms of confidence interval estimation.

\begin{sidewaystable}[htp]
	\caption{Estimates, standard errors and p-values of the linear mediation models. }
	\label{table 9}
	\begin{threeparttable}
		\resizebox{1.0\columnwidth}{!}{
			\begin{tabular}{*{7}{c}}
				\toprule
				\multirow{2}{*}{}
				& & \multicolumn{4}{c}{Renewable Method} 
				& \multicolumn{1}{c}{Full Data Method}
				\\
				\cmidrule(r){3-6} &
				& $k=10$ & $k=50$    &  $k=100$ &   $k=500$ &  $k=1$\\
				\midrule

& \multicolumn{1}{l}{$\hat{\alpha_1} \hat{\beta_1} $}
  & -1.6731 $\times 10^{-4}$
  &	-1.6731 $\times 10^{-4}$
  &-1.6731 $\times 10^{-4}$
  & -1.6731  $\times 10^{-4}$
  & -1.6731  $\times 10^{-4}$\\
 
&\multicolumn{1}{l}{$\hat{\sigma} $}
 & 3.3156 $\times 10^{-5} $
 & 3.3156 $\times 10^{-5} $
 & 3.3156 $\times 10^{-5} $
 & 3.3156 $\times 10^{-5} $
 & 3.3156 $\times 10^{-5} $\\
 
&\multicolumn{1}{l}{$P_{sobel}$ }
& $4.5048 \times 10^{-7}$ 
&  $4.5048 \times 10^{-7}$ 
& $4.5048 \times 10^{-7}$ 
& $4.5048 \times 10^{-7}$ 
& $4.5048 \times 10^{-7}$ \\

$\alpha_1 \beta_1 $
& \multicolumn{1}{l}{$P_{Asobel} $ }
& $<10^{-5}$ &$<10^{-5}$ & $<10^{-5}$ & $<10^{-5}$ & $<10^{-5}$\\

&\multicolumn{1}{l}{$P_{JS} $}   
& 4.3424  $\times 10^{-7}$
& 4.3424  $\times 10^{-7}$
& 4.3424  $\times 10^{-7}$
& 4.3424  $\times 10^{-7}$
 &4.3424 $\times 10^{-7}$\\
&\multicolumn{1}{l}{$P_{AJS} $}  
&  $<10^{-7}$
& $<10^{-7}$
& $<10^{-7}$ 
& $<10^{-7}$  
& $<10^{-7}$\\

&\multicolumn{1}{l}{$CI_{Sobel} $} 
&[-2.3230 $\times 10^{-4}$, -1.0233 $\times 10^{-4}$] 
& [-2.3230 $\times 10^{-4}$, -1.0233 $\times 10^{-4}$] 
& [-2.3230 $\times 10^{-4}$, -1.0233 $\times 10^{-4}$] 
& [-2.3230 $\times 10^{-4}$, -1.0233 $\times 10^{-4}$] 
& [-2.3230 $\times 10^{-4}$, -1.0233 $\times 10^{-4}$]\\

&\multicolumn{1}{l}{$CI_{Asobel}$ }
&[-1.9980$\times 10^{-4}$, -1.3482$\times 10^{-4}$]  
& [-1.9980$\times 10^{-4}$, -1.3482$\times 10^{-4}$]  
& [-1.9980$\times 10^{-4}$, -1.3482$\times 10^{-4}$]  
& [-1.9980$\times 10^{-4}$, -1.3482$\times 10^{-4}$]
&[-1.9980$\times 10^{-4}$, -1.3482$\times 10^{-4}$]  \\
				\midrule
				
	&\multicolumn{1}{l}{$\hat{\alpha_2} \hat{\beta_2} $ }
& 5.9063 $\times 10^{-4}$
& 5.9063  $\times 10^{-4}$
& 5.9063 $\times 10^{-4}$
& 5.9063  $\times 10^{-4}$
& 5.9063 $\times 10^{-4}$\\
	
& \multicolumn{1}{l}{$\hat{\sigma}$} 
&  6.0683 $\times 10^{-5}$
& 6.0683  $\times 10^{-5}$
& 6.0683  $\times 10^{-5}$
& 6.0683 $\times 10^{-5}$
& 6.0683  $\times 10^{-5}$\\
&\multicolumn{1}{l}{$P_{sobel}$} & $<10^{-5}$ &  $<10^{-5}$ & $<10^{-5}$ & $<10^{-5}$ & $<10^{-5}$ \\

$\alpha_2 \beta_2 $& \multicolumn{1}{l}{$P_{Asobel} $ }& $<10^{-5}$ & $<10^{-5}$ & $<10^{-5}$ &$<10^{-5}$ & $<10^{-5}$\\

& \multicolumn{1}{l}{$P_{JS} $ }
& $<10^{-7}$
& $<10^{-7}$
& $<10^{-7}$
& $<10^{-7}$
& $<10^{-7}$\\
& \multicolumn{1}{l}{$P_{AJS} $ } 
& $<10^{-7}$
& $<10^{-7}$
& $<10^{-7}$
& $<10^{-7}$
& $<10^{-7}$\\

& \multicolumn{1}{l}{$CI_{sobel}$ }
&[-7.0957$\times 10^{-4}$, -4.7169$\times 10^{-4}$] 
& [-7.0957$\times 10^{-4}$, -4.7169$\times 10^{-4}$] 
& [-7.0957$\times 10^{-4}$, -4.7169$\times 10^{-4}$] 
& [-7.0957$\times 10^{-4}$, -4.7169$\times 10^{-4}$] 
& [-7.0957$\times 10^{-4}$, -4.7169$\times 10^{-4}$]\\
& \multicolumn{1}{l}{$CI_{Asobel}$} 
&[-6.5010 $\times 10^{-4}$, -5.3116$\times 10^{-4}$]  
& [-6.5010$\times 10^{-4}$, -5.3116$\times 10^{-4}$] 
& [-6.5010$\times 10^{-4}$, -5.3116$\times 10^{-4}$] 
& [-6.5010$\times 10^{-4}$, -5.3116$\times 10^{-4}$] 
& [-6.5010$\times 10^{-4}$, -5.3116$\times 10^{-4}$]\\
\midrule
CPU time & & 0.8204 & 0.7505 &	0.8895 & 1.0933 & 5.1637 \\
	\bottomrule      

\end{tabular} }

\end{threeparttable}
\end{sidewaystable}

\begin{sidewaystable}[htp]
	\caption{Estimates, standard errors and p-values of the logistic mediation model. }
	\label{table 10}
	\begin{threeparttable}
		\resizebox{1.0\columnwidth}{!}{
			\begin{tabular}{*{7}{c}}
				\toprule
				\multirow{2}{*}{}
				& & \multicolumn{4}{c}{Renewable Method} 
				& \multicolumn{1}{c}{Full Data Method}
				\\
				\cmidrule(r){3-6} &
				& $k=10$ & $k=50$    &  $k=100$ &   $k=500$ &  $k=1$\\
				\midrule
& \multicolumn{1}{l}{$\hat{\alpha_1} \hat{\beta_1} $} & 1.0327 $\times 10^{-2}$
&	1.0329 $\times 10^{-2}$
&	1.0328 	$\times 10^{-2}$
& 1.0230  $\times 10^{-2}$
& 1.0350 $\times 10^{-2}$
\\
		
& \multicolumn{1}{l}{$\hat{\sigma}  \times 10^{3}$ }
& 2.6413  $\times 10^{-3}$
& 2.6398 $\times 10^{-3}$
& 2.6397 $\times 10^{-3}$
& 2.6362 $\times 10^{-3}$
& 2.6386 $\times 10^{-3}$ \\
& \multicolumn{1}{l}{$P_{sobel} $	}
&9.2420 $\times 10^{-5}$
&9.1209$\times 10^{-5}$
&9.1259$\times 10^{-5}$
& 9.3415$\times 10^{-5}$
& 8.7590$\times 10^{-5}$	\\
$\alpha_1 \beta_{1}$	& \multicolumn{1}{l}{$P_{Asobel} $ }
&9.2420$\times 10^{-5}$
&9.1209 $\times 10^{-5}$
&9.1259$\times 10^{-5}$
& 9.3415$\times 10^{-5}$
& 8.7590$\times 10^{-5}$\\
& \multicolumn{1}{l}{$P_{JS} $ }
&9.2312 $\times 10^{-5}$
&9.1102$\times 10^{-5}$
&9.1152$\times 10^{-5}$
& 9.3306$\times 10^{-5}$
& 8.7486 $\times 10^{-5}$\\
& \multicolumn{1}{l}{$P_{AJS}$ }
&9.2312 $\times 10^{-5}$
&9.1102 $\times 10^{-5}$
&9.1152$\times 10^{-5}$
&9.3306$\times 10^{-5}$
& 8.7486$\times 10^{-5}$\\	

&\multicolumn{1}{l}{$CI_{sobel}$	}
&[0.5150 $ \times 10^{-2}$,1.5504 $ \times 10^{-2}$] 
&[0.5155$ \times 10^{-2}$,1.5503$ \times 10^{-2}$]
&[0.5155$ \times 10^{-2}$,1.5502$ \times 10^{-2}$]
&[0.5133$ \times 10^{-2}$,1.5467$ \times 10^{-2}$]
& [0.5179$ \times 10^{-2}$,1.5522$ \times 10^{-2}$]	\\
& \multicolumn{1}{l}{$CI_{Asobel} $ }
&[0.5150$ \times 10^{-2}$,1.5504$ \times 10^{-2}$]
&[0.5155$ \times 10^{-2}$,1.5503$ \times 10^{-2}$]
&[0.5155$ \times 10^{-2}$,1.5502$ \times 10^{-2}$]
&[0.5133$ \times 10^{-2}$,1.5467$ \times 10^{-2}$]
& [0.7765$ \times 10^{-2}$,1.2936$ \times 10^{-2}$]\\
\midrule

&\multicolumn{1}{l}{$\hat{\alpha_2} \hat{\beta_2} $ }& -0.1045
&	-0.1045
& -0.1045
&	-0.1045 
& -0.1045 \\

& \multicolumn{1}{l}{$\hat{\sigma} $ }
& 1.2817  $\times 10^{-3}$
&	1.2815  $\times 10^{-3}$
&	1.2815  $\times 10^{-3}$
&	1.2813  $\times 10^{-3}$
&1.2818  $\times 10^{-3}$
\\
& \multicolumn{1}{l}{$P_{sobel} $}	&$< 10^{-5}$ &$< 10^{-5}$ &$< 10^{-5}$ &$< 10^{-5}$ &	$< 10^{-5}$\\
$\alpha_2 \beta_{2}$ & \multicolumn{1}{l}{$P_{Asobel} $} &$< 10^{-5}$ &$< 10^{-5}$ &$< 10^{-5}$ &$< 10^{-5}$ & $< 10^{-5}$\\
& \multicolumn{1}{l}{$P_{JS} $} &$< 10^{-5}$ &$< 10^{-5}$ &$< 10^{-5}$ & $< 10^{-5}$ & $< 10^{-5}$\\
& \multicolumn{1}{l}{$P_{AJS} $ }&$< 10^{-5}$ &$< 10^{-5}$ &$< 10^{-5}$ & $< 10^{-5}$ & $< 10^{-5}$\\
&\multicolumn{1}{l}{	$CI_{sobel} $}	&[-0.1070,-0.1020]&[-0.1070, -0.1020] & [ -0.1070,-0.1020] &[-0.1070,-0.1020] &[-0.1070,-0.1020]	\\
& \multicolumn{1}{l}{$CI_{Asobel} $} &[-0.1070,-0.1020]&[-0.1070, -0.1020] &[-0.1070,-0.1020] &[-0.1070,-0.1020]& [-0.1059,-0.1033]\\
\midrule
				CPU time & &2.4119 & 2.5578 &	0.78912 & 1.1612 & 7.3322 \\
				\bottomrule      
				
		\end{tabular} }
		
	\end{threeparttable}
\end{sidewaystable}

\subsection{Binary Outcomes}
The renewable method is employed in this section to analyze a large dataset on the loan data of Lending Club from 2007 to 2015, which is publicly available at \url{https:// www.kaggle.com/wendykan/lending-club-data}. According to \cite{Kumar2007}, there exists a significant association between the total loan amount and both the borrower's annual income and loan status.  The risk of defaulting is higher for individuals who have taken out larger loans. Additionally, a borrower's annual income is expected to be associated with their interest rate, and those with higher rates may encounter difficulties in meeting repayment deadlines \citep{2015DeterminantsOD}.
Our objective is to examine whether the loan amount and interest rate act as mediators in the relationship between annual income and default status.
 The exposure $X$ is defined as the borrower's annual income, measured in units of 10,000. The binary variable $Y$ represents the borrower's loan status, where $Y = 1$ indicates default and $Y = 0$ signifies punctual repayment. Two mediator variables, namely $M_1$ and $M_2$, denote the total loan amount (expressed in ten thousand dollars) and the interest rate of the loan respectively. The repayment period, denoted as a covariate $Z_1$, was categorized into two levels: $Z_1 = 1$ for a duration of 36 months and $Z_1 = 2$ for a duration of 60 months. To facilitate analysis, the continuous variables $X$, $M_1$, and $M_2$ were standardized with zero mean and unit variance, respectively. 
The complete dataset was obtained by excluding missing data, with a sample size of $N_k$ = 825,994.

The logistic mediation models fitted with the complete dataset are presented as follows:
\begin{align*}
P(Y =1 )&=\dfrac{\exp(-4.260 - 0.367 X + 0.044 M_1 + 2.225 M_2 - 0.323 Z_1)}
	{1 + \exp(-4.260 - 0.367 X + 0.044 M_1 + 2.225 M_2 - 0.323 Z_1)},
\\
M_1 &= 0.146 + 0.234 X + 0.418 Z_1 +  e_1,
\\
 M_2 &= 0.594 - 0.047 X + 0.299 Z_1 +  e_2.
\end{align*}
The full data is divided into $k$ blocks, denoted as $\boldsymbol{D_{1}}, \ldots, \boldsymbol{D_{k}}$, with each block having a sample size of $n_k=N_k/k$. We use our proposed online updating methods for mediation analysis in streaming data. The value of $k$ is chosen as 10, 50, 100, and 500 respectively.
The estimates, standard errors, p-values, and 95\% confidence intervals of mediation effects using the renewable method and the full data method are presented in Table \ref{table 10}. Additionally, the computational efficiency is evaluated by utilizing the provided CPU time (in seconds). The results in Table \ref{table 10} indicate that the renewable approach consistently produces comparable estimates to the full data method across different batch sizes, while significantly improving computational efficiency.

\section{Concluding Remarks}
The present study examined the statistical inference of mediation effects with streaming data in the context of linear and logistic mediation models. The study introduced four mediation testing methods, namely the Sobel test, ASobel test, JS test, and AJS test. Additionally, it provided the confidence interval of mediation effects based on streaming data for both the Sobel and ASobel methods. The effectiveness of our renewable method in practical applications was demonstrated through simulations and two real-world examples.

There exist two topics to be studied in the future. First, the conventional mediation methods assume that all confounders can be measured, which is often unverifiable in the case of large datasets. In certain scenarios, such as streaming data, confounders may exhibit time-varying characteristics \citep{time_varying_confounders}. It is desirable to extend the proposed method to accommodate situations where time-varying confounders exist.
Second, the application of survival analysis in the context of big data has gained significant popularity \cite[]{JCGS-Cox}. It is intriguing to explore the integration of streaming survival data with mediation analysis.

\section{Appendix}
\setcounter{equation}{0}
In the Appendix, we provide the details of the derivation for $\tilde{\phi}^{(k)}$ as given in equation \eqref{A1}. Note that
\begin{align}
\tilde{\phi}^{(k)} & = \dfrac{1}{N_k - (1+p+q)} \sum\limits_{i=1}^{k} \big(\boldsymbol{Y}^{(i)} - \boldsymbol{W}^{(i)}  \boldsymbol{\tilde{\Gamma}}^{(k)}  \big)^{\bm{'}} 
\big(\boldsymbol{Y}^{(i)}  - \boldsymbol{W}^{(i)}  \boldsymbol{\tilde{\Gamma}}^{(k)}  \big) \nonumber \\
& = \dfrac{1}{N_k - (1+p+q)} \sum\limits_{i=1}^{k}
\big(
(\boldsymbol{Y}^{(i)})^{'} \boldsymbol{Y}^{(i)} - (\boldsymbol{\tilde{\Gamma}}^{(k)})^{'} (\boldsymbol{W}^{(i)} )^{'}\boldsymbol{Y}^{(i)} -
(\boldsymbol{Y}^{(i)})^{'} \boldsymbol{W}^{(i)} \boldsymbol{\tilde{\Gamma}}^{(k)} \nonumber
\\&+ (\boldsymbol{\tilde{\Gamma}}^{(k)})^{'} (\boldsymbol{W}^{(i)} )^{'}
\boldsymbol{W}^{(i)} \boldsymbol{\tilde{\Gamma}}^{(k)}
\big) \nonumber \nonumber \\
& = \dfrac{1}{N_k - (1+p+q)} 
\Big\{
\sum\limits_{i=1}^{k-1} (\boldsymbol{Y}^{(i)})^{'} \boldsymbol{Y}^{(i)} + (\boldsymbol{Y}^{(k)})^{'} \boldsymbol{Y}^{(k)}
 -2 \sum\limits_{i=1}^{k-1}  (\boldsymbol{\tilde{\Gamma}}^{(k)})^{'} (\boldsymbol{W}^{(i)} )^{'} \boldsymbol{Y}^{(i)} \nonumber \\
&- (\boldsymbol{\tilde{\Gamma}}^{(k)})^{'} (\boldsymbol{W}^{(k)} )^{'} \boldsymbol{Y}^{(k)}
- (\boldsymbol{Y}^{(k)})^{'} \boldsymbol{W}^{(k)} \boldsymbol{\tilde{\Gamma}}^{(k)} 
+
(\boldsymbol{\tilde{\Gamma}}^{(k)})^{'}
\boldsymbol{\tilde{J}}^{(k)}  \boldsymbol{\tilde{\Gamma}}^{(k)} 
\Big\}.
\label{EQ-7-18}
\end{align}
In view of the fact that
\begin{align*}
\boldsymbol{\tilde{\Gamma }}^{(k)} &= 
\big( \boldsymbol{\tilde{J}}^{(k-1)}  + \boldsymbol{J}^{(k)} \big)^{-1} 
\big( \boldsymbol{\tilde{J}}^{(k-1)} \boldsymbol{\tilde{\Gamma }}^{(k-1)} + (\boldsymbol{W}^{(k)})^{\boldsymbol{'}} \boldsymbol{Y}^{(k)} \big),
\end{align*}
then we have 
\begin{align}
(\boldsymbol{W}^{(k)})^{\boldsymbol{'}} \boldsymbol{Y}^{(k)} = \boldsymbol{\tilde{J}}^{(k)} \boldsymbol{\tilde{\Gamma }}^{(k)} - \boldsymbol{\tilde{J}}^{(k-1)} \boldsymbol{\tilde{\Gamma }}^{(k-1)}.
\label{EQ-7-19}
\end{align}
Thus, \eqref{EQ-7-18} can be written as
\begin{align}
\tilde{\phi}^{(k)} & = \dfrac{1}{N_k - (1+p+q)} 
\big\{
\sum\limits_{i=1}^{k-1} (\boldsymbol{Y}^{(i)})^{'} \boldsymbol{Y}^{(i)} + (\boldsymbol{Y}^{(k)})^{'} \boldsymbol{Y}^{(k)}
-2 \sum\limits_{i=1}^{k-1}  (\boldsymbol{\tilde{\Gamma}}^{(k)})^{'} (\boldsymbol{W}^{(i)} )^{'} \boldsymbol{Y}^{(i)} \nonumber \\
& -(\boldsymbol{\tilde{\Gamma}}^{(k)})^{'} (\boldsymbol{\tilde{J}}^{(k)} \boldsymbol{\tilde{\Gamma}}^{(k)} - \boldsymbol{\tilde{J}}^{(k-1)} \boldsymbol{\tilde{\Gamma}}^{(k-1)})
- (\boldsymbol{\tilde{J}}^{(k)} \boldsymbol{\tilde{\Gamma}}^{(k)} - \boldsymbol{\tilde{J}}^{(k-1)} \boldsymbol{\tilde{\Gamma}}^{(k-1)})^{'} \boldsymbol{\tilde{\Gamma}}^{(k)}
+
(\boldsymbol{\tilde{\Gamma}}^{(k)})^{'}
\boldsymbol{\tilde{J}}^{(k)}  \boldsymbol{\tilde{\Gamma}}^{(k)} 
\big\} \nonumber \\
&= \dfrac{1}{N_k - (1+p+q)} 
\Big\{
\sum\limits_{i=1}^{k-1} (\boldsymbol{Y}^{(i)})^{'} \boldsymbol{Y}^{(i)} + (\boldsymbol{Y}^{(k)})^{'} \boldsymbol{Y}^{(k)}
-2 \sum\limits_{i=1}^{k-1}  (\boldsymbol{\tilde{\Gamma}}^{(k)})^{'} (\boldsymbol{W}^{(i)} )^{'} \boldsymbol{Y}^{(i)} - (\boldsymbol{\tilde{\Gamma}}^{(k)})^{'} \boldsymbol{\tilde{J}}^{(k)} \boldsymbol{\tilde{\Gamma}}^{(k)} \nonumber \\
&
+ (\boldsymbol{\tilde{\Gamma}}^{(k)})^{'} \boldsymbol{\tilde{J}}^{(k-1)} \boldsymbol{\tilde{\Gamma}}^{(k-1)}
- (\boldsymbol{\tilde{\Gamma}}^{(k)})^{'}
(\boldsymbol{\tilde{J}}^{(k)})^{'} 
\boldsymbol{\tilde{\Gamma}}^{(k)}
+ (\boldsymbol{\tilde{\Gamma}}^{(k-1)})^{'}
(\boldsymbol{\tilde{J}}^{(k-1)})^{'} 
\boldsymbol{\tilde{\Gamma}}^{(k)}
+(\boldsymbol{\tilde{\Gamma}}^{(k)})^{'} \boldsymbol{\tilde{J}}^{(k)} \boldsymbol{\tilde{\Gamma}}^{(k)}
\Big\} \nonumber\\
&=  \dfrac{1}{N_k - (1+p+q)} 
\Big\{
\sum\limits_{i=1}^{k-1} (\boldsymbol{Y}^{(i)})^{'} \boldsymbol{Y}^{(i)} + (\boldsymbol{Y}^{(k)})^{'} \boldsymbol{Y}^{(k)}
-2 \sum\limits_{i=1}^{k-1}  (\boldsymbol{\tilde{\Gamma}}^{(k)})^{'} (\boldsymbol{W}^{(i)} )^{'} \boldsymbol{Y}^{(i)} \nonumber \\
& - (\boldsymbol{\tilde{\Gamma}}^{(k)})^{'} \boldsymbol{\tilde{J}}^{(k)} \boldsymbol{\tilde{\Gamma}}^{(k)} + 
2 (\boldsymbol{\tilde{\Gamma}}^{(k)})^{'} \boldsymbol{\tilde{J}}^{(k-1)} \boldsymbol{\tilde{\Gamma}}^{(k-1)}
\Big\}.  \label{EQ-7-20}
\end{align}
It is straightforward to prove that
\begin{eqnarray}
-2 \sum\limits_{i=1}^{k-1}  (\boldsymbol{\tilde{\Gamma}}^{(k)})^{'} (\boldsymbol{W}^{(i)} )^{'} \boldsymbol{Y}^{(i)}
+ 2 (\boldsymbol{\tilde{\Gamma}}^{(k)})^{'} \boldsymbol{\tilde{J}}^{(k-1)} \boldsymbol{\tilde{\Gamma}}^{(k-1)}
= 
-2 \sum\limits_{i=1}^{k-1}  (\boldsymbol{\tilde{\Gamma}}^{(k-1)})^{'} (\boldsymbol{W}^{(i)} )^{'} \boldsymbol{Y}^{(i)}
+ 2 (\boldsymbol{\tilde{\Gamma}}^{(k-1)})^{'} \boldsymbol{\tilde{J}}^{(k-1)} \boldsymbol{\tilde{\Gamma}}^{(k-1)}.
\label{EQ-7-21}
\end{eqnarray}
From \eqref{EQ-7-19}, we can get that
\begin{align*}
(\boldsymbol{W}^{(i)})^{\boldsymbol{'}} \boldsymbol{Y}^{(i)} = \boldsymbol{\tilde{J}}^{(i)} \boldsymbol{\tilde{\Gamma }}^{(i)} - \boldsymbol{\tilde{J}}^{(i-1)} \boldsymbol{\tilde{\Gamma }}^{(i-1)}, ~ i=1,\ldots,k.
\end{align*}
Accordingly,  the left part of \eqref{EQ-7-21} can be calculated as follows,
\begin{align*}
&-2 \sum\limits_{i=1}^{k-1}  (\boldsymbol{\tilde{\Gamma}}^{(k)})^{'} (\boldsymbol{W}^{(i)} )^{'} \boldsymbol{Y}^{(i)}
+ 2 (\boldsymbol{\tilde{\Gamma}}^{(k)})^{'} \boldsymbol{\tilde{J}}^{(k-1)} \boldsymbol{\tilde{\Gamma}}^{(k-1)}\\
&= -2(\boldsymbol{\tilde{\Gamma}}^{(k)})^{'}
[(\boldsymbol{W}^{(1)} )^{'} \boldsymbol{Y}^{(1)}
+ 
(\boldsymbol{W}^{(2)} )^{'} \boldsymbol{Y}^{(2)}
+
\cdots
+
(\boldsymbol{W}^{(k-1)} )^{'} \boldsymbol{Y}^{(k-1)}]
+ 2 (\boldsymbol{\tilde{\Gamma}}^{(k)})^{'} \boldsymbol{\tilde{J}}^{(k-1)} \boldsymbol{\tilde{\Gamma}}^{(k-1)} \\
&= -2(\boldsymbol{\tilde{\Gamma}}^{(k)})^{'}
[\boldsymbol{\tilde{J}}^{(1)} \boldsymbol{\tilde{\Gamma }}^{(1)} - \boldsymbol{\tilde{J}}^{(0)} \boldsymbol{\tilde{\Gamma }}^{(0)}
+
\boldsymbol{\tilde{J}}^{(2)} \boldsymbol{\tilde{\Gamma }}^{(2)} - \boldsymbol{\tilde{J}}^{(1)} \boldsymbol{\tilde{\Gamma }}^{(1)}
+
\cdots
+
\boldsymbol{\tilde{J}}^{(k-1)} \boldsymbol{\tilde{\Gamma }}^{(k-1)} - \boldsymbol{\tilde{J}}^{(k-2)} \boldsymbol{\tilde{\Gamma }}^{(k-2)}]
\\&+
2 (\boldsymbol{\tilde{\Gamma}}^{(k)})^{'} \boldsymbol{\tilde{J}}^{(k-1)} \boldsymbol{\tilde{\Gamma}}^{(k-1)}\\
&= -2(\boldsymbol{\tilde{\Gamma}}^{(k)})^{'} 
[\boldsymbol{\tilde{J}}^{(k-1)} \boldsymbol{\tilde{\Gamma }}^{(k-1)} -
\boldsymbol{\tilde{J}}^{(0)} \boldsymbol{\tilde{\Gamma }}^{(0)}
]
+
2 (\boldsymbol{\tilde{\Gamma}}^{(k)})^{'} \boldsymbol{\tilde{J}}^{(k-1)} \boldsymbol{\tilde{\Gamma}}^{(k-1)}.
\end{align*}
Due to $\boldsymbol{\tilde{J}}^{(0)} = \boldsymbol{0}$ and $\boldsymbol{\tilde{\Gamma }}^{(0)} = \boldsymbol{0}$, we observe that the left part of \eqref{EQ-7-21} is
$$
-2 \sum\limits_{i=1}^{k-1}  (\boldsymbol{\tilde{\Gamma}}^{(k)})^{'} (\boldsymbol{W}^{(i)} )^{'} \boldsymbol{Y}^{(i)}
+ 2 (\boldsymbol{\tilde{\Gamma}}^{(k)})^{'} \boldsymbol{\tilde{J}}^{(k-1)} \boldsymbol{\tilde{\Gamma}}^{(k-1)} =0.
$$
Similary, the right part of \eqref{EQ-7-21} satisfies
$$
-2 \sum\limits_{i=1}^{k-1}  (\boldsymbol{\tilde{\Gamma}}^{(k-1)})^{'} (\boldsymbol{W}^{(i)} )^{'} \boldsymbol{Y}^{(i)}
+ 2 (\boldsymbol{\tilde{\Gamma}}^{(k-1)})^{'} \boldsymbol{\tilde{J}}^{(k-1)} \boldsymbol{\tilde{\Gamma}}^{(k-1)} =0.
$$
Based on \eqref{EQ-7-21} and \eqref{EQ-7-20}, we have 
\begin{align*}
\tilde{\phi}^{(k)} & = \dfrac{1}{N_k - (1+p+q)}
\Big\{
\sum\limits_{i=1}^{k-1} (\boldsymbol{Y}^{(i)})^{'} \boldsymbol{Y}^{(i)} + (\boldsymbol{Y}^{(k)})^{'} \boldsymbol{Y}^{(k)}
-2 \sum\limits_{i=1}^{k-1}  (\boldsymbol{\tilde{\Gamma}}^{(k-1)})^{'} (\boldsymbol{W}^{(i)} )^{'} \boldsymbol{Y}^{(i)} \nonumber \\
& - (\boldsymbol{\tilde{\Gamma}}^{(k)})^{'} \boldsymbol{\tilde{J}}^{(k)} \boldsymbol{\tilde{\Gamma}}^{(k)} + 
2 (\boldsymbol{\tilde{\Gamma}}^{(k-1)})^{'} \boldsymbol{\tilde{J}}^{(k-1)} \boldsymbol{\tilde{\Gamma}}^{(k-1)}
\Big\}\\
& = \dfrac{1}{N_k - (1+p+q)}
\Big\{
\sum\limits_{i=1}^{k-1} (\boldsymbol{Y}^{(i)} - \boldsymbol{W}^{(i)}  \boldsymbol{\tilde{\Gamma}}^{(k-1)}  )^{\bm{'}} 
(\boldsymbol{Y}^{(i)}  - \boldsymbol{W}^{(i)}  \boldsymbol{\tilde{\Gamma}}^{(k-1)} )  \nonumber
+ (\boldsymbol{\tilde{\Gamma}}^{(k-1)})^{\bm{'}} \boldsymbol{\tilde{J}}^{(k-1)}  \boldsymbol{\tilde{\Gamma}}^{(k-1)} \\
&+ 
(\boldsymbol{Y}^{(k)}){'}\boldsymbol{ Y}^{(k)}  -(\boldsymbol{\tilde{\Gamma}}^{(k)})^{\bm{'}} \boldsymbol{\tilde{J}}^{(k)} \boldsymbol{\tilde{\Gamma}}^{(k)}
\Big\} \\
& = \dfrac{1}{N_k - (1+p+q)} \Big\{ \big(N_{k-1} - (1+p+q)  \big)\tilde{\phi}^{(k-1)}  + (\boldsymbol{\tilde{\Gamma}}^{(k-1)})^{\bm{'}} \boldsymbol{\tilde{J}}^{(k-1)}  \boldsymbol{\tilde{\Gamma}}^{(k-1)}  \\
& + 
(\boldsymbol{Y}^{(k)}){'}\boldsymbol{ Y}^{(k)}  -(\boldsymbol{\tilde{\Gamma}}^{(k)})^{\bm{'}} \boldsymbol{\tilde{J}}^{(k)} \boldsymbol{\tilde{\Gamma}}^{(k)}  \Big\} .
\end{align*}
\bibliographystyle{apalike}
\bibliography{reference2}

\end{document}